\documentclass[twocolumn]{aa}
\usepackage{txfonts,textcomp,natbib,amssymb,graphicx}
\begin{document}
\title{Near-Infrared long-slit spectra of Seyfert Galaxies: gas excitation across the central kiloparsec}

\author{T.P.R. van der Laan \inst{1,2}
\and
E. Schinnerer \inst{2}
\and
T. B\"oker \inst{3}
\and
L. Armus \inst{4}}

\institute{Institute de Radioastronomie Millimetrique (IRAM), 300 Rue de la Piscine, 38406 St. Martin d'Heres, Grenoble, France\\ \email{vanderlaan@iram.fr}
\and
Max-Planck-Institut f\"ur Astronomie, K\"onigstuhl 17, 69117 Heidelberg, Germany
\and
European Space Agency, Keplerlaan 1, 2200AG Noordwijk, The Netherlands
\and
Spitzer Science Center, California Institute of Technology, Pasadena, CA 91125, USA}

\date{Received 22 August 2013; Accepted 11 October 2013}

\abstract
{The excitation of the gas phase of the interstellar medium can be driven by various mechanisms. In galaxies with an active nucleus, such as Seyfert galaxies, both radiative and mechanical energy from the central black hole, or the stars in the disk surrounding it may play a role.}{We investigate the relative importance and range of influence of the active galactic nucleus for the excitation of ionized and molecular gas in the central kiloparsec of its host galaxy.}{We present H- and K-band long-slit spectra for a sample of 21 nearby (D $<$70\,Mpc) Seyfert galaxies obtained with the NIRSPEC instrument on the Keck telescope. For each galaxy, we fit the nebular line emission, stellar continua, and warm molecular gas as a function of distance from the nucleus.}
{Our analysis does not reveal a clear difference between the nucleus proper and off-nuclear environment in terms of excitation mechanisms, suggesting that the influence of an AGN reaches far into the disk of the host galaxy. The radial variations in emission line ratios indicate that, while local mechanisms 
do affect the gas excitation, they are often averaged out when measuring over extended regions. The fully calibrated long-slit spectra, as well as our fitting results, are made available on-line.}{}

\titlerunning{Gas excitation across the central kiloparsec}

\maketitle

\section{Introduction}
\footnotetext{The data presented herein were obtained at the W.M. Keck Observatory, which is operated as a scientific partnership among the California Institute of Technology, the University of California and the National Aeronautics and Space Administration. The Observatory was made possible by the generous financial support of the W.M. Keck Foundation. Long-slit data are available in electronic form at the CDS via anonymous ftp to cdsarc.u-strasbg.fr (130.79.128.5) or via http://cdsweb.u-strasbg.fr/cgi-bin/qcat?J/A+A/560/A99}
The physical processes that govern the excitation of gas in galaxies have been the subject of intense study for many years. Especially in the centers of galaxies with active galactic nuclei (AGN), more than one process may contribute to the observed line emission. Atoms may be photo-ionized by stellar UV-photons (UV-fluorescence), or heated via young stars, or via X-rays emitted by the AGN. Alternatively, mechanical shocks from either supernovae or an outflow driven by the AGN may play a role. In principle, photo-ionization, thermal heating, and mechanical shocks affect the interstellar medium (ISM) differently, and a comparison of the relative strength and distribution of various emission lines should therefore be able to distinguish between excitation mechanisms and sources \citep[e.g.][]{Black1987,1989ApJ...342..306H,1993ApJ...416..150F,Simpson1996}. 

The well-known BPT diagram \citep{BPT1981} distinguishes between starburst (SB), AGN or LINER excitation using the optical emission lines, [OIII], [NII], H$\alpha$, and H$\beta$. In the near-IR there is no similar diagram, but this wavelength range does have several other advantages over the optical to determine the dominant excitation mechanism. 

Warm H$_2$ emission has several strong vibrational transitions in this region. The combination of multiple H$_2$ lines makes it possible to distinguish between non-thermal (SB via UV-fluorescence) or thermal (SB, AGN, shocks) mechanisms \citep[e.g.][]{Black1987,1989ApJ...342..306H,Maloney1996}. In the case of UV-fluorescence, H$_2$ molecules are excited via absorption of photons in the Lyman-Werner bands in the near-UV. Such photons are generated in young stars, and diffuse into the ISM over significant distances, and are thus not coupled to the local temperature (i.e. non-thermal). By comparison, any number of sources may interact directly with the local ISM, and thus affect the local temperature (i.e. thermal). Which process cannot be distinguished by H$_2$ lines alone.

A second advantage is that key [FeII] emission lines are present at 1.26$\mu$m (J-band) and 1.64\,$\mu$m (H-band). Strong [FeII] emission is only possible when dust grains are destroyed, which frees iron atoms into the ISM \citep{1993ApJ...416..150F,Simpson1996}. Dust grain destruction occurs when shocks or an AGN are present \citep{Mouri2000}. 

Finally, ionizing radiation can be traced through the Pa$\beta$ and Br$\gamma$ transitions of hydrogen at 1.28\,$\mu$m and 2.17\,$\mu$m, respectively. Therefore, a comparison of H$_2$ and [FeII] emission line strengths to those of hydrogen recombination lines provide an indicator of the dominant excitation source \citep{Larkin1998,Rodriguez2004,Rodriguez2005}, similar to the BPT diagram in the optical regime.

This NIR diagnostic is most often utilized with the ratios [FeII]1.26$\mu$m/Pa$\alpha$ and H$_2$/Br$\gamma$, since this minimizes relative reddening effects on the emission lines. A tight correlation between these ratios was found, see \citet{2013MNRAS.430.2002R} for the most up-to-date version, based on a sample of 65 spectra of star forming, AGN, and LINER galaxy nuclei. 

Recent photoionization models have shed light on this correlation. \citet{2012MNRAS.422..252D} presented (non-spatial) photoionization models where the ionizing source was composed of two components, an AGN emitting X-rays, and cosmic rays. With these models the authors were able to reproduce line ratio observations, recover the correlation, and show that the [FeII]1.26$\mu$m/Pa$\alpha$ ratio is predominantly driven by gas-phase metallicity and the H$_2$/Br$\gamma$ ratio by the strength of the X-ray emission. Weakening the X-ray contribution moved the models from the observed domain. The models presented in \citet{2013MNRAS.430.2002R} focus more the H$_2$ heating by either AGN of SB. These authors argue that, while AGN X-ray excitation may play an important role, multiple mechanisms will affect the gas excitation depending on the level of nuclear activity.

Thus, at any position in the circumnuclear region the excitation of gas will be a combination, depending on the local physical excitation conditions there. Quantifying the radial variations in gas excitation tracers thus provides vital clues to which excitation source dominates as a function of radius.

In this paper, we present near-infrared, long-slit H- and K-band spectra for a sample of 21 nearby Seyfert galaxies, to examine this question. The paper is structured as follows. The sample selection, observations, and data reduction, as well as the derivation of emission line fluxes, are summarized in \S2. In \S3 we investigate the excitation mechanisms, based on [FeII], Br$\gamma$ and H$_2$ emission line fluxes in the full long-slit spectra, and in a comparison of those lines in the nuclear and off-nuclear environments. In \S4 the kinematics of the lines are presented and used to place further constraints on the excitation mechanisms. \S5 contains the discussion, \S6 the summary. Finally, the appendix holds an overview of the data products that are made public with this work. The long-slit spectra presented here are very rich, and contain much more information than is discussed, such as additional emission lines, and the stellar continuum. 

\section{Observations and data reduction}
\subsection{Sample selection}
The sample of Seyfert galaxies used for this study was selected from the \citet{Whittle1992} and the \citet{Ho1997} catalogs based on their H$\beta$ and [OIII] $\lambda$ 5007\AA{} line emission. More specifically, we required that F(H$\beta$) $>$ 10$^{-14}$ erg s$^{-1}$ cm$^{-2}$ and F([OIII]) $>$ 10$^{-13}$ erg s$^{-1}$ cm$^{-2}$ in order to provide high SNR NIR spectra, see also \citet{1995ApJ...448...98H} and \citet{Gonzalez2001}.

To achieve the highest spatial resolution, the sample was limited to galaxies with a recession velocity below 5000\,km\,s$^{-1}$. Given the typical NIR seeing at Mauna Kea of better than 0.7\arcsec\,, a spatial resolution element then generally corresponds to less than 250\,pc.

Finally, for the targets to be well observable from Mauna Kea (+20\degr\,latitude), a declination limit of -25\degr\, was imposed. The final sample meeting these requirements consists of 33 Seyfert galaxies. Due to time allocation and weather constraints, 21 targets were observed, as summarized in Table\,\ref{tab:sample}. The final sample contains 10 Seyfert 1's and 11 Seyfert 2's, based on the classification of \citet{2006A&A...455..773V}.  

\begin{table*}
\caption{Observed sample, with observation log and Seyfert type.}\label{tab:sample}
\centering
\begin{tabular}{l c c c r c c c c c l}
\hline\hline
Source & RA (J2000) & Dec (J2000) & Type & V$_{LSR}$ & D [Mpc] & t$_{int}$ (s) & airmass & Seeing [\arcsec] & PA slit & Calibrator star \\
\hline
Mrk 1		&01h16m07.2s &  33d05m22s &S2   & 4824 & 61.4 & 1200/1800 & 1.24 & 0.70 & 266 & BD+42384 \\
Mrk 3		&06h15m36.3s &  71d02m15s &S1h  & 3998 & 55.2 & 1200 	& 1.58 & 0.75 &  30 & HD\,23650 \\
Mrk 1066		&02h59m58.6s &  36d49m14s &S2   & 3565 & 46 .7 & 1800 	& 1.01 & 0.50 & -90 & HD\,19061 \\
NGC 0262	&00h48m47.1s &  31d57m25s &S1h  & 4507 & 57.4 & 1800 	& 1.10 &  0.60 & 267 & BD+42384 \\
NGC 0931	&02h28m14.5s &  31d18m42s &S1   & 4989 & 65.1 & 1800 	& 1.19 & 0.50 & 255 & HD\,19061 \\
NGC 1320	&03h24m48.7s & -03d02m33s &S2   & 2663 & 34.2 & 1200 	& 1.18 & 0.70 & 145 & HD\,32665 \\
NGC 1667	&04h48m37.1s & -06d19m12s &S2   & 4574 & 61.7 & 1800/1500 & 1.49 & 0.70 & 200 & HD\,36869 \\
NGC 2110	&05h52m11.4s & -07d27m22s &S1i  & 2335 & 32.9 & 1800 	& 1.08 & 0.65 & 161 & HD\,292795 \\
NGC 2273	&06h50m08.7s &  60d50m45s &S1h  & 1840 & 25.7 & 1800 	& 1.19 & 1.15 & 50 & HD\,90733 \\
NGC 2685	&08h55m34.6s &  58d44m03s &S2   & 883  & 13.7 & 1200 	& 1.26 & 0.70 & 36 & HD\,72911 \\
NGC 2992	&09h45m42.0s & -14d19m35s &S1i  & 2314 & 36.3 & 1200 	& 1.20 & 0.85 & 195 & HD\,85538 \\
NGC 3081	&09h59m29.5s & -22d49m35s &S1h  & 2391 & 37.4 & 1800 	& 1.49 & 0.70 & 158 & HD\,66325 \\
NGC 3185	&10h17m38.5s &  21d41m18s &S2   & 1230 & 21.0 & 1800 	& 1.21 & 0.55 & -45 & HD\,95364 \\
NGC 3516	&11h06m47.5s &  72d34m07s &S1.5 & 2593 & 37.2 & 1800 	& 1.56 & 0.55 & -120 & HD\,124446 \\
NGC 4051	&12h03m09.6s &  44d31m53s &S1n  & 725  & 12.7 & 1800 	& 1.11 & 0.75 & -45 & SAO44110 \\
NGC 4151	&12h10m32.6s &  39d24m21s &S1.5 & 995  & 17.0 & 1200/1800 & 1.05 & 0.80 & 335 & HD\,128611 \\
NGC 4253	&12h18m26.5s &  29d48m46s &S1n  & 3876 & 56.9 & 1200 	& 1.05 & 0.45 & 290 & HD\,129290 \\
NGC 4388	&12h25m46.7s &  12d39m44s &S1h  & 2517 & 39.1 & 1200 	& 1.21 & 0.65 & 272 & HD\,121876 \\
NGC 5273	&13h42m08.3s &  35d39m15s &S1.9 & 1022 & 17.5 & 2700 	& 1.15 & 1.00 & 370 & HD\,129290 \\
NGC 5506	&14h13m14.9s & -03d12m27s &S1i  & 1824 & 28.9 & 1200 	& 1.23 & 0.80 & 90 & HD\,131428 \\
NGC 7450	&23h00m47.8s & -12d55m07s &S1.5 & 3185 & 38.8 & 1800 	& 1.30 & 0.65 & 35 & HD\,5106\\
\hline
\end{tabular}
\tablefoot{Columns 1-3: source name and J2000 coordinates; 
Column 4: Seyfert type from \citet{2006A&A...455..773V}; (S1h: S2 which show S1 like spectra in polarized light, S1i: S1 with a broad Paschen H$\beta$ line, S1n: narrow-line Seyfert 1). 
Column 5: heliocentric velocity (from NED); 
Column 6: heliocentric radial distance (NED); 
Column 7: on-source integration time (when not equal in H- and K-bands, H-band is listed first); 
Column 8: mean airmass during the integrations. 
Column 9: mean seeing, i.e. FWHM of the PSF as measured from the calibrator star; 
Column 10: P.A. of NIRSPEC slit (N to E); 
Column 11: ID of calibrator star }
\end{table*}

\subsection{Observations}\label{sec:obs}
\begin{figure*}
\centering
\includegraphics[width=16.8cm]{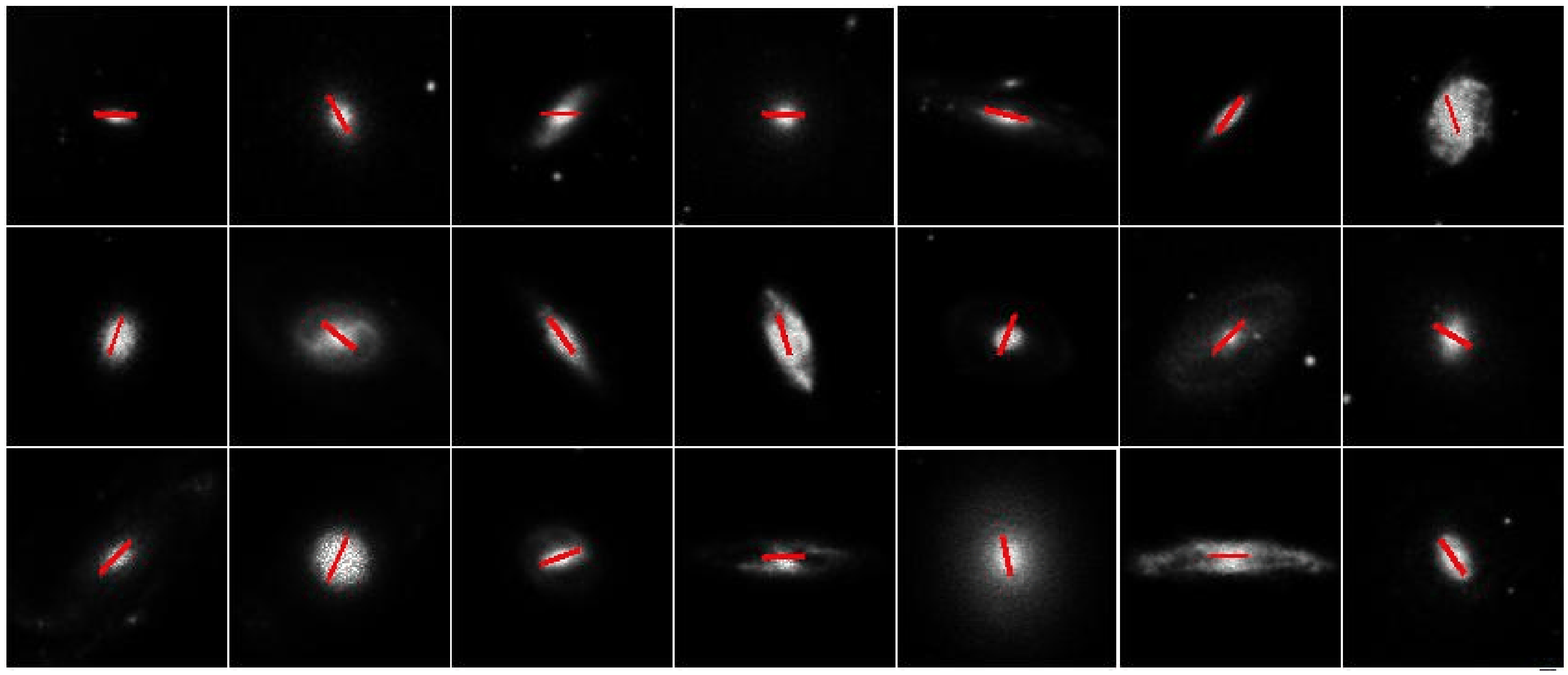} 
\caption{Mosaic of all 21 galaxies in our sample with the NIRSPEC 21\arcsec\, useable slit superimposed. Positioning (from left-to-right, top-to-bottom) follows Table \ref{tab:sample}. Images taken from the DSS, with a FOV of 2\arcmin\,. The images have been flux-scaled to show the approximate extent observable with the NIRSPEC pre-imaging option, as used to target each source during the NIRSPEC long-slit observations. {\it A color version of this figure is available in the online journal.}}\label{fig:targets}
\end{figure*}

All 21 targets were observed in the H (1.5 - 1.8 $\mu\textrm{m}$) and K (2.0 - 2.4 $\mu$m) band, using the Keck NIR longslit spectrograph NIRSPEC \citep{1998SPIE.3354..566M}. The observations were performed during two observing runs in 2001, March 11-12, and 2001, November, 12-13. The average seeing in the NIR was 0.65 - 0.70\arcsec , as measured from the calibration stars, except for the first night which had a seeing of about 0.90\arcsec . A slit length of $0.772\arcsec \times 42\arcsec$ with a pixel scale of 0.144\arcsec /pixel was used. The corresponding spectral resolution in the H- and K-band are $\lambda/\Delta\lambda = 1700$ and $\lambda/\Delta\lambda = 1570$, respectively. The slit orientation was selected to follow the (optical) major axis of each target (Fig. \ref{fig:targets}). The galaxy nucleus was positioned in the slit using the pre-imaging option of the NIRSPEC camera, such that along-slit nodding with an amplitude of half the slit length (i.e. leaving a useable 21\arcsec\, wide spectrum) could be accommodated.

The integration time per exposure was 5\,min, split in 4 frames of 75\,s each. Integrations were typically 20-30\,min per band, only NGC\,5273 was observed for 45\,min. Except for one source (NGC\,5506), this set-up avoided strong non-linearity and/or saturation of the detector. In NGC\,5506, the spectra of the very nucleus (i.e. inner 4 pixels) were saturated, but the rest of the spectrum is fully usable.

For flux calibration and atmosphere correction, a nearby standard star of type G2V was observed immediately before and/or after the target observations. In addition, internal lamp spectra were obtained in each band and for each target configuration. Finally, sky flats were taken at the beginning and/or end of each observing night.

The final spectral resolution of the spectra is 2.86\AA\, per spectral pixel in the H-band and 4.27\AA\, per spectral pixel in the K-band.

\subsection{Data Reduction}\label{sec:datareduc}
Data reduction was done in IRAF \citep{1993ASPC...52..173T}. We followed standard procedure and performed i) spatial and spectral flat-fielding using sky flats and darks, ii) wavelength calibration using identified lines in the lamp spectra as well as strong atmospheric OH lines and iii) distortion correction in the spectral domain using atmospheric OH lines that are present in each frame.

The telluric absorption features and instrument response were corrected together as follows. Each observed standard star spectrum (the G2 dwarfs) was compared to a synthetic spectrum for the same stellar type. The synthetic spectra were generated at the same spectral resolution as the data from the models of \citet{Ariane}. Dividing the observed stellar spectrum by the synthetic one then yields a ``correction spectrum'' containing both the telluric absorption and the instrument response. The observed long-slit galaxy spectra were divided by this correction spectrum. The division was done separately for the H- and K-band. In general, the corrected galaxy spectra are free of telluric features. In NGC\,0262 some artifacts in the CO bandhead wavelength range remain. Further visual inspection showed that some sky lines remain visible in the spectra of NGC\,4051 and NGC\,5273, but for our purposes this can be accounted for in the later fitting procedure and has no detrimental effect on our results.

Finally, each galaxy spectrum was flux-calibrated using the H- and K-band magnitude of the standard stars as given in the 2MASS catalog \citep{2MASS2003}. The flux calibration relies on the assumption that the conversion factor derived for the star is also valid for the corresponding galaxy observation. While changes in sky transparency or airmass throughout the night can add uncertainty to the conversion, each galaxy-star pair was observed closely together in time and at similar airmass.

In a few cases the narrow slit width of 0.772\arcsec\, is smaller than the FWHM of the (seeing-limited) stellar PSF. This can result in different flux losses for star and galaxy, and thus introduce errors in the {\it absolute} calibration of the galaxy spectra. Moreover, the stellar PSF is not identical in the H- and K-band spectra, possibly causing {\it relative} calibration errors between both bands.

Several checks were preformed to gauge the quality and consistency of the flux calibration. First, the final combined H+K stellar spectra were visibly inspected for a smooth, continuous spectral distribution. This comparison demonstrated that the calibrated spectra can be reasonably fit by single black-body curve, confirming that the relative flux calibration is correct and robust.

It is difficult to verify absolute flux measurements in Seyfert galaxies, because their continuum emission can vary over time by up to a factor of 2-3 \citep[i.e.][]{Glass2004,Breedt2010,Sakata2010}. A comparison between our results and other works (Table \ref{tab:fluxvalues}), shows that our flux measurements agree to first order. The studies referenced in Table \ref{tab:fluxvalues} \citep{Rodriguez2004,Rodriguez2005,Ramos2009} use spectra extracted over broader slit apertures (1.5\arcsec - 2\arcsec), 
and we have scaled their results to match ours (0.772\arcsec $\times$ 0.144\arcsec, central pixel-position). Overall, we consider the agreement
to be reasonable.

\begin{table}
\caption{Comparison of absolute flux calibration}\label{tab:fluxvalues}
\centering
\begin{tabular}{l | c | c c}
\hline\hline
Object & Our work & \multicolumn{2}{c}{Other work}\\
 & $10^{-11}$\,erg\,s$^{-1}$\,cm$^{-2}$\,$\mu$m$^{-1}$ & \multicolumn{2}{c}{$10^{-11}$\,erg\,s$^{-1}$\,cm$^{-2}$\,$\mu$m$^{-1}$}\\
\hline
Mrk\,1066 & 0.3 & 0.1 (1) & 0.5 (2)\\
NGC\,0262 & 0.4 & 0.06 (1) & 0.11 (2)\\
NGC\,4051 & 0.6 & 0.2 (3) & 1.6 (4)\\
\hline
\end{tabular}
\tablefoot{Comparison of continuum flux density between our work and published values. Values listed are in H-band for Mrk\,1066 and NGC\,0262, and in K-band for NGC\,4051.}
\tablebib{(1) \citet{Rodriguez2005}; (2) \citet{Ramos2009}; (3) \citet{Rodriguez2004}; (4) \citet{2008MNRAS.385.1129R}}
\end{table}

\subsection{Fitting stellar and gaseous components}\label{sec:GANDALF}
The pPXF \citep{CappEms04} and GANDALF fitting software\footnote{The pPXF and GANDALF routines can be downloaded from 
http://www-astro.physics.ox.ac.uk/$\sim$mxc/idl/ and http://star-www.herts.ac.uk/$\sim$sarzi/PaperV\_nutshell/PaperV\_nutshell.html, respectively.} \citep{Sarzi2006,Jesus2006} was used to fit the stellar and gaseous emission in each spectrum along the slit. The pPXF code deals more specifically with the stellar continuum and the GANDALF code more with the emission lines. The fitting was performed separately in H- and K-band for each position along the slit. The stellar continuum was fitted with a weighted combination of stellar spectra, taken from a template library. Unfortunately, all empirical NIR stellar spectral libraries available in the literature have lower spectral resolution than offered by the NIRSPEC observations. In order to not degrade the spectral resolution of the data, a synthetic spectral library based on the models of \citet{Ariane} was kindly provided to us by A. Lancon and subsequently used. As has been shown by \citet{Ariane}, the library spectra can well reproduce the observed spectra of giants and supergiants in the near-infrared. As giant and supergiant stars dominate the NIR stellar emission from galaxy disks, the library should therefore be well suited for our purposes.

The emission lines were fitted with Gaussian profiles according to a list of prominent emission lines in the H- and K-band wavelength range (Table \ref{tab:lines}). These emission lines include Br$\gamma$, [FeII] at 1.64$\mu\textrm{m}$, multiple rotational and vibrational lines of H$_{2}$, and two coronal lines (CaVIII and NeII). An emission line is considered `detected' if the amplitude of the Gaussian exceeds the 3$\sigma$ noise level. 

\begin{table}
\caption{Emission lines included in the spectral fitting}\label{tab:lines}
\centering
\begin{tabular}{l c | l c}
\hline\hline
Line & $\lambda$ [$\mu\textrm{m}$] & Line & $\lambda$ [$\mu\textrm{m}$] \\
\hline
$[$FeII$]$ ($a^4F_{9/2}-a^4D_{5/2}$)	& 1.5330 	& NeII ($^2P_{3/2}-^2Po_{1/2} $)				& 2.1480 \\
$[$FeII$]$ ($a^4F_{9/2}-a^4D_{7/2}$)	& 1.6440 	& H$_{2}$(2-1)S(2)		& 2.1542 \\
$[$FeII$]$ ($a^4F_{7/2}-a^4D_{5/2}$)& 1.6790 	& Br$\gamma$			& 2.1661 \\
Br$\zeta$ 			& 1.7360 	& H$_{2}$(3-2)S(3)		& 2.2014 \\
H$_{2}$(1-0)S(7)		& 1.7470	& H$_{2}$(1-0)S(0)		& 2.2233 \\
H$_{2}$(1-0)S(2)		& 2.0338 	& H$_{2}$(2-1)S(1)		& 2.2477 \\
HeI $^1P-^1S$			& 2.0587 	& CaVIII ($^2Po_{1/2}-^2Po_{3/2}$)& 2.3211 \\
H$_{2}$(3-2)S(5)		& 2.0656 	& H$_{2}$(2-1)S(0)		& 2.3550 \\
H$_{2}$(2-1)S(3)		& 2.0737 	& H$_{2}$(3-2)S(1)		& 2.3858 \\
H$_{2}$(1-0)S(1)		& 2.1218 	&  & \\
\hline
\end{tabular}
\end{table}

As starting values for the fit, we used the V$_{LSR}$ values of Table \ref{tab:sample} for the stellar and emission line velocity, and an initial emission line width of 50\,km\,s$^{-1}$. The emission line velocities were constrained to be within 500\,km\,s$^{-1}$ from the starting value, which allows for both overall galactic rotation as well as possible radial motions, without leading to confusion between different emission lines. The velocity dispersion was not constrained, since gas in various environments (NLR, BLR, and/or the galaxy disk) may be probed.

The pPXF + GANDALF code accounts for the instrumental resolution ($\sim$190\,km\,s$^{-1}$ in H-band, $\sim$176\,km\,s$^{-1}$ in K band). The fitted emission lines were allowed to have individual velocities and dispersions, except for the H$_{2}$ lines, which were assumed to originate from the same gas clouds, and thus should have identical kinematics. (Inspection of the fitting results confirms this assumption.) More specifically, we forced GANDALF to apply the velocity and velocity dispersion of the strongest H$_{2}$ line, the (1-0)S(1) transition at 2.1218$\mu\textrm{m}$, to all other H$_{2}$ lines. 

\begin{figure*}
\centering
\includegraphics{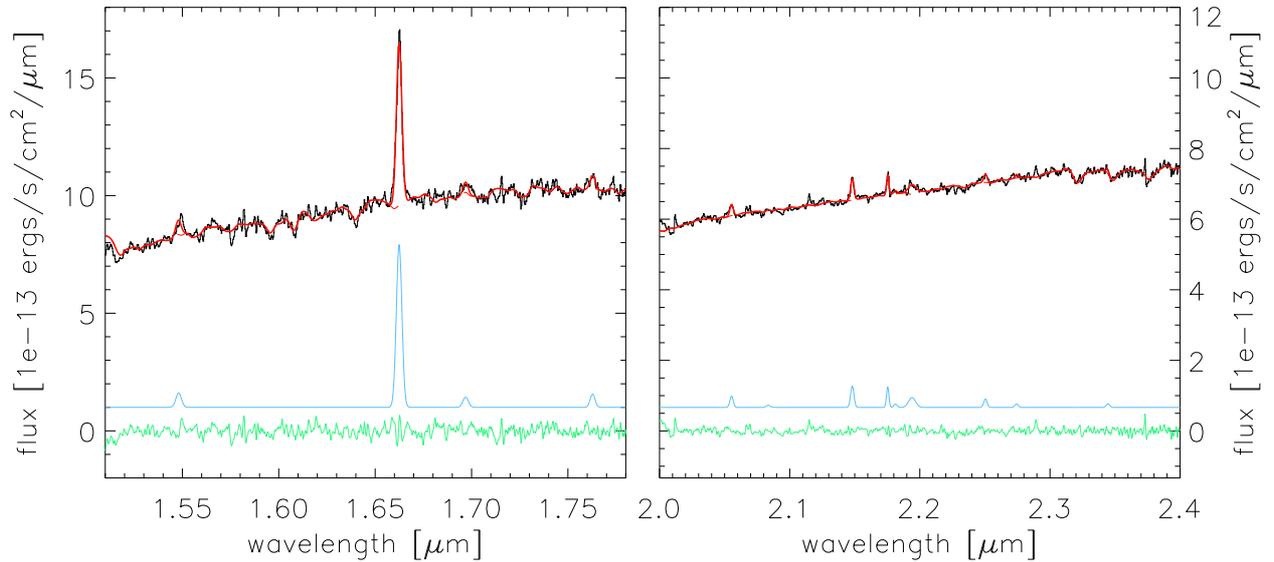}
\caption{An example of the spectral fitting with the pPXF and GANDALF routines. These H- (left) and K-band (right) spectra are from the nucleus of NGC\,2110 (0.65\arcsec\, aperture). The black line represents the observed spectrum and the red line the best-fit returned by pPXF + GANDALF. This fit includes both the composite stellar spectrum and emission lines. The green line shows the residual (observation - fit). The blue line shows the combined emission line fit. This line has been offset from zero for better display. Shown is one spatial position, each spatial position in the H- and K-band spectra was fitted independently. {\it A color version of this figure is available in the online journal.}}
\label{fig:GANDALFfit}
\end{figure*}

An example of the resulting fit is presented in Fig. \ref{fig:GANDALFfit}. The black line shows the observed spectrum, while the red line shows the best fit, i.e. the sum of the best fitting stellar template mix and the various detected emission lines. In addition, we show the ``pure'' derived emission line spectrum (blue) and the fit residuals (green).

Our observations reach high continuum signal-to-noise levels. Therefore, the error of a measured emission line flux is dominated by the pPXF and GANDALF fitting routine, and can be recovered by the amplitude-to-noise (A/N) estimate. Here, amplitude is the central, maximum, height of the fitted gaussian, after accounting for the stellar continuum (and the possible stellar absorption lines at those wavelengths). Noise, in this sense, is the local residual of the fit (combining the stellar templates and significant emission lines). Together, the inverse A/N ratio gives a good approximation for the relative error in the emission line flux. 

The accuracy of GANDALF in recovering the kinematics of the emission lines was investigated by \citet{Sarzi2006}. The fitting uncertainty in the emission line velocity determination depends on the A/N of the line. Above an A/N of 5, this generally leads to accuracies in velocity of $\sim$20-40\,km/s. The accuracy in the velocity dispersion is less dependent on the A/N (at least at this range), and is $\sim$20\,km/s.

\subsection{Nuclear vs. off-nuclear apertures}\label{sec:nucoff}
The good seeing and large radial range of the observations allow us to study the dominant excitation mechanism as a continuous function
of distance from the nucleus. However, for an additional check later on, whether the AGN indeed has a more significant influence on the gas excitation within the central $\sim$100\,pc, we also define nuclear and off-nuclear apertures.

The radial range probed by each long-slit spectrum depends on the distance to the source, as well as the signal-to-noise achieved over the integration time. For nearby sources, the slit length limits the radial range, while for more distant sources, the flux drops below the detection limit at large radii. All galaxies are detected out to a radius of at least $\sim$350\,pc from the nucleus. The resolution limit imposed by the atmospheric seeing varies across the sample from 70 to 230\,pc. For a consistent analysis throughout the sample, we define the `nuclear' region as equal to the seeing-limited PSF appropriate for
each galaxy. For comparison, we also define two `off-nuclear' regions spanning the radial range from $\sim$125\,pc to $\sim$350\,pc,
from which we extract spectra from either side of the nucleus. The individual spectra within each aperture are integrated over radius. The resulting, aperture integrated, spectra are run through pPXF and GANDALF to obtain the line emission fluxes.

\begin{figure*}
\centering
\includegraphics[]{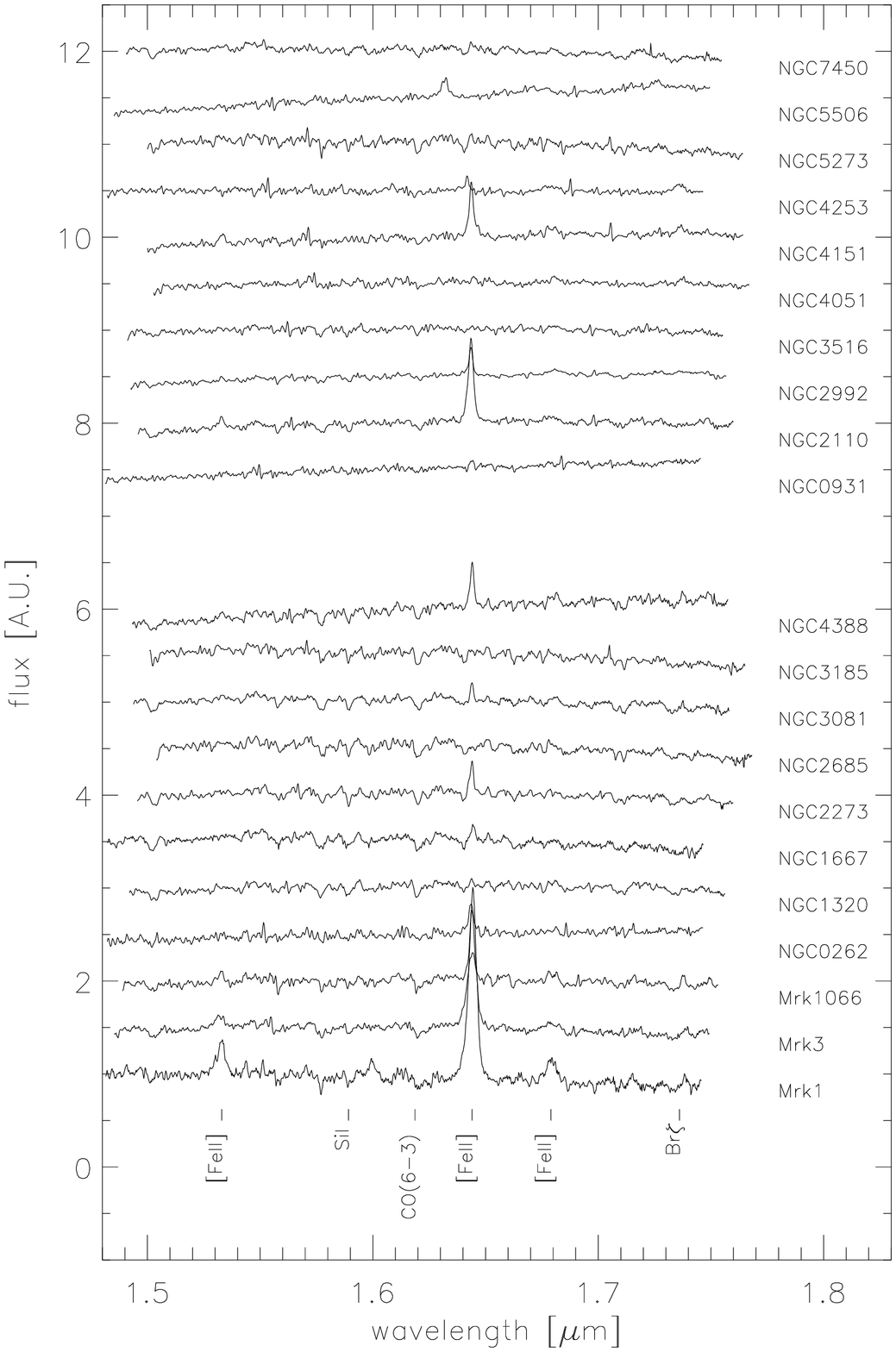}
\caption{H-band spectra of the integrated nuclear apertures for all 21 sources in the sample. Spectra are shown in their rest frame wavelengths and scaled for display. Key emission lines and absorption feature in the H-band are indicated.}\label{fig:Hband}
\end{figure*}

\begin{figure*}
\centering
\includegraphics[]{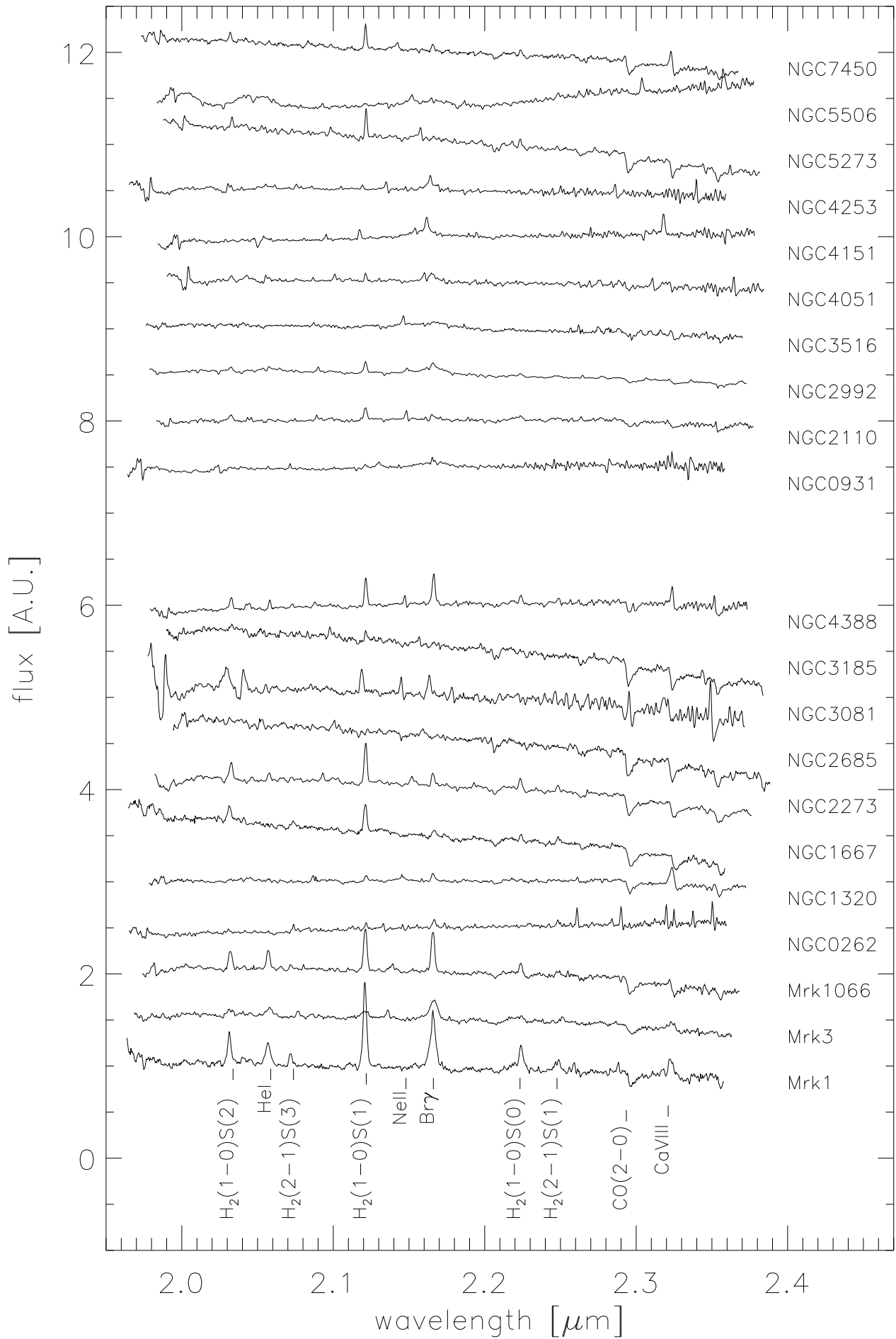}
\caption{K band spectra of the integrated nuclear apertures for all 21 sources in the sample. Spectra are shown in their rest frame wavelengths and scaled for display. Key emission lines and absorption feature in the K-band are indicated. The nuclear K band spectrum of NGC\,5506 was overexposed.}\label{fig:Kband}
\end{figure*}

The H- and K-band spectra of the nuclear cut-outs, as well as the emission line fluxes of the key lines, are presented in Figs. \ref{fig:Hband} and \ref{fig:Kband}, and Table \ref{tab:fluxes}, respectively.

\section{Categorizing gas excitation}
\subsection{Line intensities and ratios}\label{sec:ratio}
\begin{figure*}
\centering
\includegraphics[]{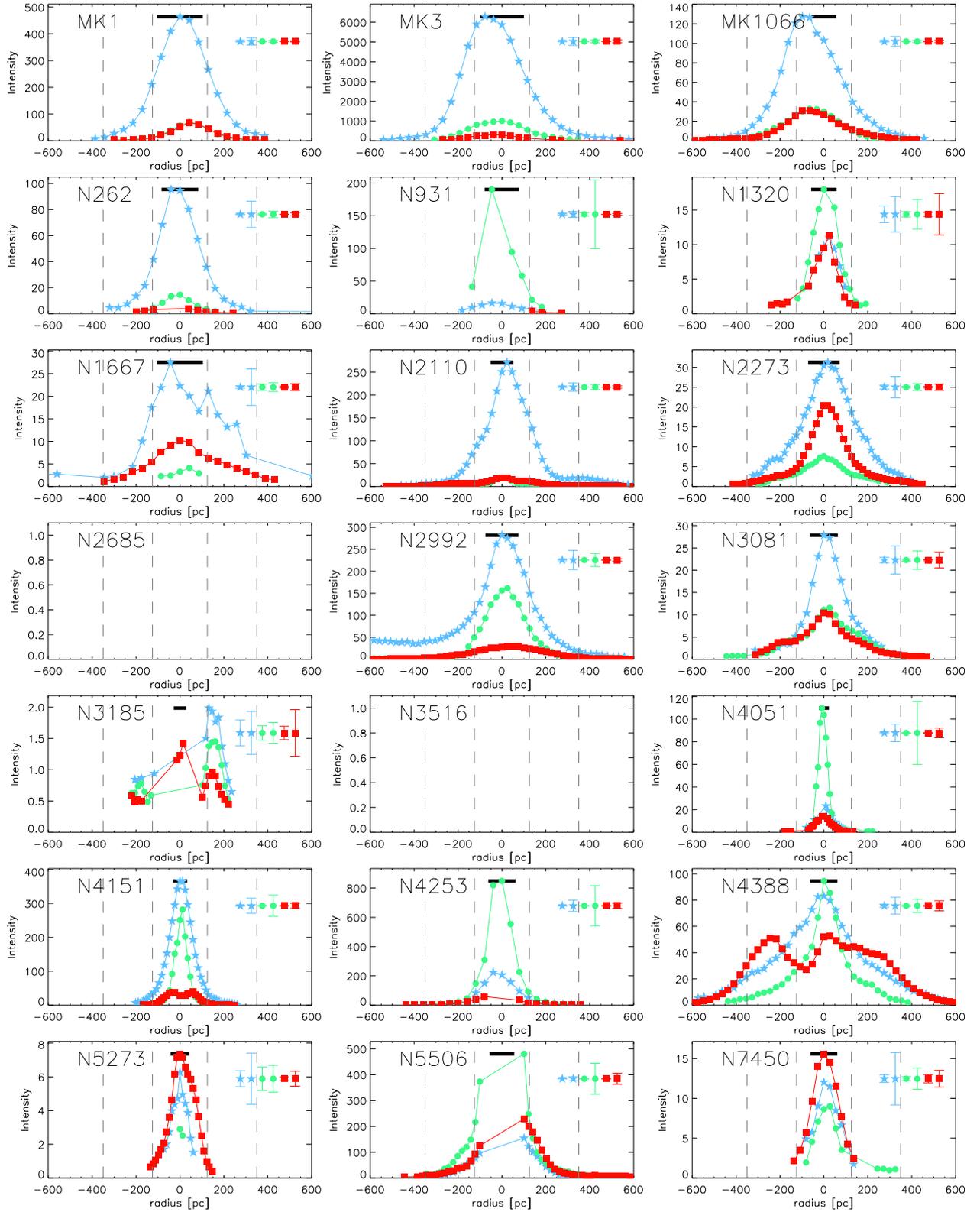}
\caption{The radial distribution of the intensities for the [FeII] (blue stars), Br$\gamma$ (green dots), and H$_2$ (red squares) emission lines as derived from GANDALF. Displayed points each have an S/N\,$>$\,3. The seeing disk (and nuclear aperture) is indicated with the black bar. The off-nuclear apertures are delimited by the vertical dashed lines at (+/-)375\,pc and (+/-)125\,pc. The minimum and maximum 1$\sigma$ error for each emission line are displayed in the upper right corner. {\it A color version of this figure is available in the online journal.}}\label{fig:mosiacint}
\end{figure*}

The radial profiles across the central 1.2\,kpc of the emission line fluxes of [FeII]\,1.64\,$\mu$m (blue), Br$\gamma$ (green), and H$_2$(1-0)s(1) (red), are plotted in Fig. \ref{fig:mosiacint}.  All three lines are detected with an A/N $>$ 3 at at least one position in each source, except in NGC\,2685 and NGC\,3516. The seeing disk is indicated by the thick black line at the top of each panel. The radial range of the off-nuclear apertures, defined in Sect. \ref{sec:nucoff}, and used in Sect. \ref{sec:nirdiag}, is indicated by the vertical dashed lines.

At first glance, the emission lines all follow the continuum emission in that their flux peaks at the center. The only exception is the H$_2$ emission in NGC\,4151 and NGC\,4388, which shows a small depression towards the nucleus. The former is consistent with a central hole in the H$_2$(1-0)s(1) distribution observed by \citet{1999MNRAS.305..319F,Storchi2009}. NGC\,3185 shows a secondary peak at the location of the circumnuclear star forming ring \citep{2010MNRAS.402.2462C}. 

\begin{figure*}
\centering
\includegraphics[]{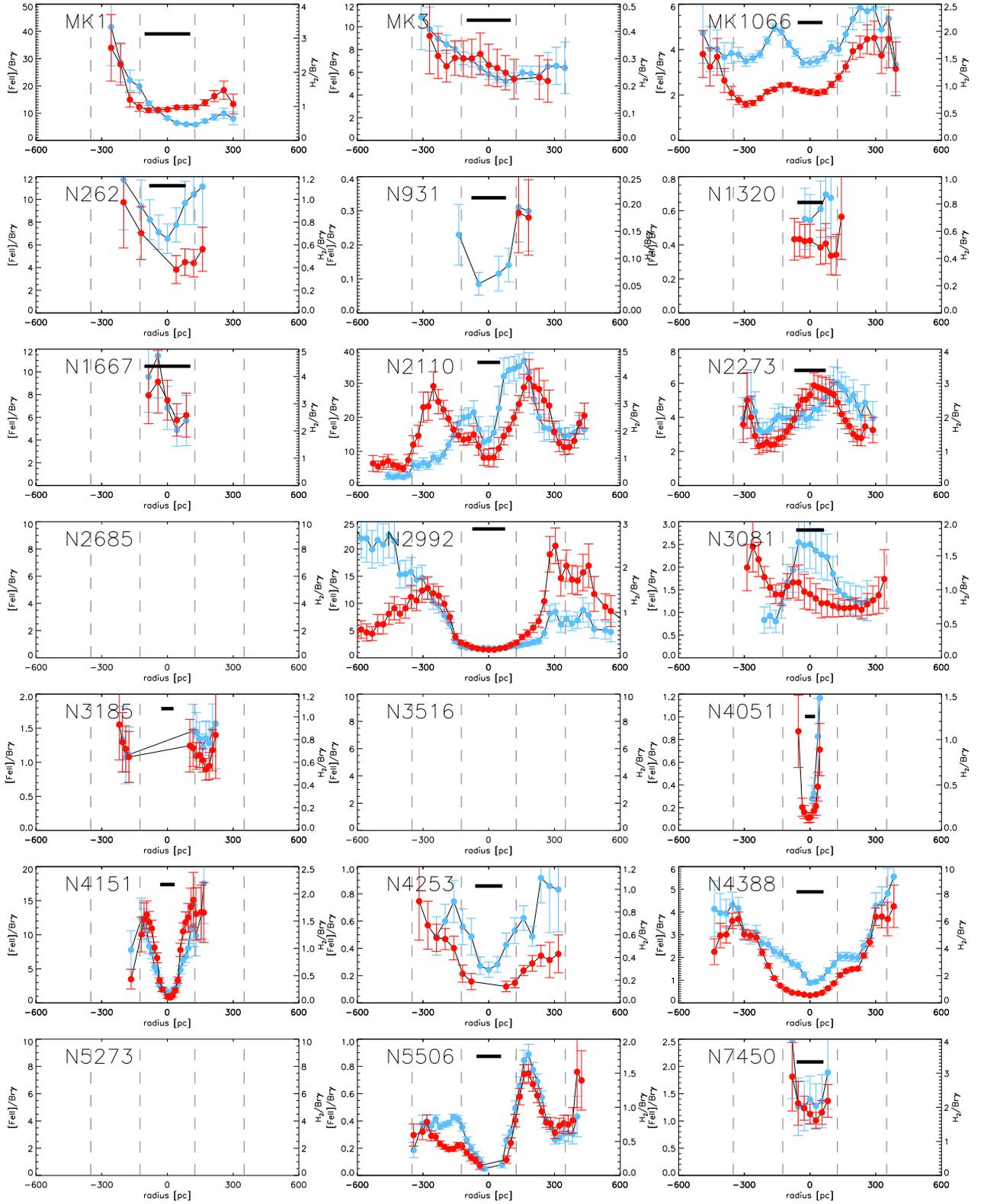}
\caption{The [FeII]/Br$\gamma$ (blue dots) and H$_2$/Br$\gamma$ (red dots) ratios, with 1$\sigma$ errors for each source. The seeing (and nuclear aperture) is indicated with the black bar. The off-nuclear apertures are delimited by the vertical dashed lines at (+/-)375\,pc and (+/-)125\,pc. {\it A color version of this figure is available in the online journal.}}\label{fig:mosiacratio}
\end{figure*}

In Fig. \ref{fig:mosiacratio}, we show the radial profile of the [FeII]/Br$\gamma$ (blue line) and H$_2$/Br$\gamma$ (red line) line ratios.  For most galaxies, both line ratios follow each other closely across the galaxy disk, which suggests that all three emission lines are produced by the same excitation sources. This is a non-trivial observation, since [FeII] and H$_2$ trace different mechanisms, e.g. mechanical (shocks) and thermal heating (radiation field), whose origin (AGN or stars) may vary independently with radius. Interestingly, the ratios drop towards the nucleus in many cases. This is in apparent contradiction to the higher values one would expect near the nucleus, given the power-law radiation and the more massive jet/outflow winds from the AGN, compared to excitation from stars in the disk \citep[e.g.][]{Black1987,Hollenbach1989,1993ApJ...416..150F,Maloney1996,Simpson1996,Mouri2000}. 

The drop in (especially) the [FeII]1.64$\mu$m/Br$\gamma$ ratio could be an effect of increased reddening towards the nucleus. However, similar drops towards the nucleus have also been observed, with high spatial resolution integral field spectrographs, in [FeII]1.26$\mu$m/Pa$\beta$ ratio maps \citep[e.g.][]{Storchi2009,2010MNRAS.404..166R,2011MNRAS.416..493R,2013MNRAS.430.2249R}. The ratio [FeII]1.26$\mu$m/Pa$\beta$ is nearly insensitive to reddening, due to the small wavelength separation between the two lines.

\begin{figure*}
\centering
\includegraphics[]{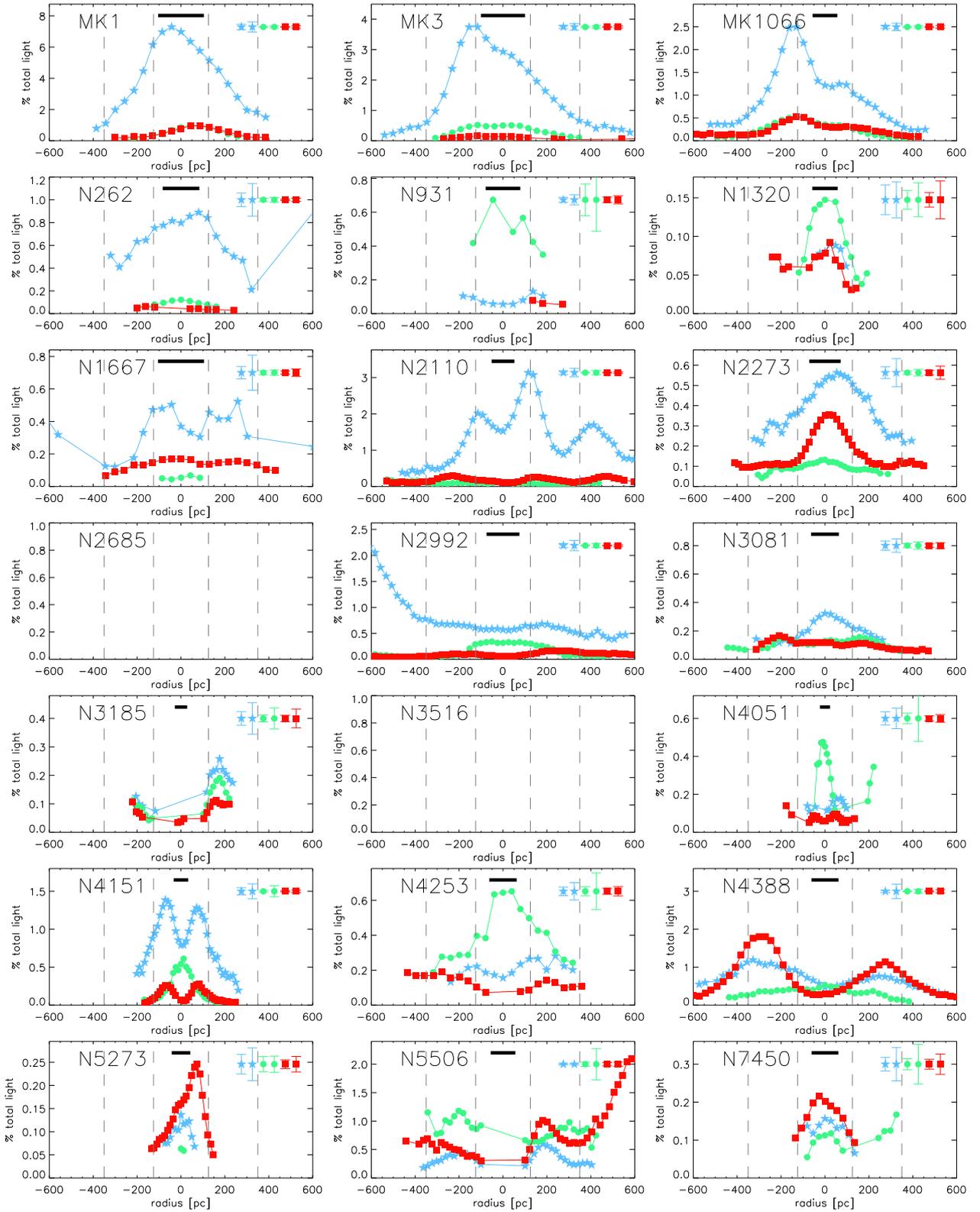}
\caption{The line-to-continuum ratios for the [FeII] (blue stars), Br$\gamma$ (green dots), and H$_2$ (red squares) emission lines. Displayed points each have an S/N\,$>$\,3. The seeing (and nuclear aperture) is indicated with the black bar. The off-nuclear apertures are delimited by the vertical dashed lines at (+/-)375\,pc and (+/-)125\,pc. The minimum and maximum 1$\sigma$ error for each emission line are displayed in the upper right corner. {\it A color version of this figure is available in the online journal.}}\label{fig:mosiacfrac}
\end{figure*}

The line-to-continuum ratios for the three emission lines are shown in Fig. \ref{fig:mosiacfrac}. This diagnostic is sensitive to gas density variations within the galaxy, since it traces second order light variations, i.e. the gas abundance in relation to the stellar light. The H$_2$ line-to-continuum ratio is constant or drops significantly towards the nucleus in most galaxies, except NGC\,2273, NGC\,5273, and NGC\,7450. For Br$\gamma$, in contrast, the ratio peaks towards the nucleus in many galaxies, consistent with stronger ionization mechanisms towards the active nucleus. These two results both explain the observed central drop of the H$_2$/Br$\gamma$ ratio.

Of the three emission lines, the [FeII] line-to-continuum ratio shows the most distinct radial variations. Its profile is centrally peaked in Mrk\,1, Mrk\,3, NGC\,262, NGC\,1320, NGC\,2273, NGC\,3081, NGC\,5273, and NGC\,7450. Several sources show two maxima (or a central drop, e.g. NGC\,1667, NGC\,3185, NGC\,4151, NGC\,4253, NGC\,4388, and NGC\,5506). Finally, Mrk\,1066, NGC\,931, NGC\,2110, NGC\,2992, and NGC\,4051 show peculiar distributions, e.g. the distribution in NGC\,2110 shows three separate peaks. 

The variation in the line-to-continuum profile of [FeII] suggest that the mechanical heating in Seyfert nuclei is more varied than the thermal heating. In galaxy nuclei where the [FeII] line-to-continuum ratio is peaked, either the AGN more effectively destroys dust grains and, consequently, liberates Fe atoms, or there is a large reservoir of dust, or both. A central drop in the [FeII]/Br$\gamma$ ratio, on the other hand, could imply that the ionizing radiation is either so strong that it has further ionized the iron atoms, or that the AGN is currently not effective in destroying dust grains.

We find that the `[FeII] peak' group contains mostly Seyfert type 2 nuclei, indicating that the peak may be linked to a larger dust reservoir (as expected in the standard AGN paradigm), or a more effective mechanism to destroy the dust and liberate the iron atoms. The peculiar group consists almost exclusively of Seyfert type 1 sources, where we are more likely to observe the nucleus though its BLR or NLR cone, or close to a jet outflow.

\subsection{Nuclear/off-nuclear comparison}\label{sec:nirdiag} 
\citet{Larkin1998} and \citet{Rodriguez2004,Rodriguez2005} proposed the line ratios [FeII]1.26$\mu$m/Pa$\beta$ and H$_2$/Br$\gamma$ as a useful diagnostic to distinguish between SB, AGN, and LINER excitation in galactic nuclei. Higher values in both ratios were shown to correlate with AGN and LINER nuclei, while lower values were consistent with star formation being
the dominant excitation mechanism. Here, we investigate the extent of the AGN influence by adding to previous studies data points for the off-nuclear apertures defined in Sect. \ref{sec:nucoff}.

\begin{figure*}
\includegraphics{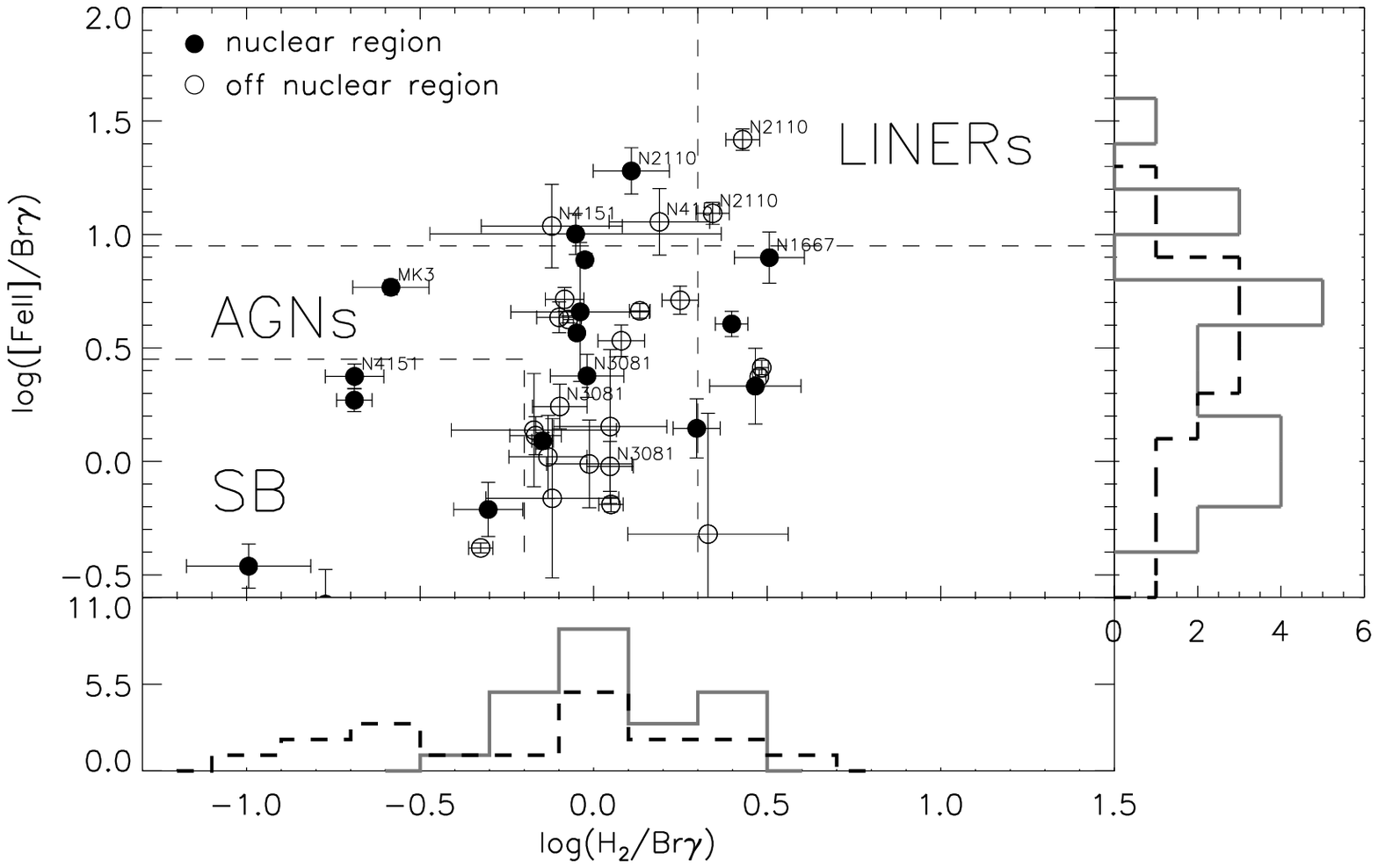}
\caption{NIR diagnostic diagram, adapted from \citet{Larkin1998} and \citet{Rodriguez2004,Rodriguez2005}. The different domains of the plot, labelled `SB' (starburst), `AGN' and `LINERs' have been shown to provide a basic classification of galactic nuclei.  We only plot the result for those apertures where we had significant line detections in all three lines. The side panels show the distributions of the two ratios, differentiated by nuclear (black, dashed) and off-nuclear (grey, solid) environment.}
\label{fig:BPT}
\end{figure*}

The spectral range of our data does not contain the [FeII]1.26$\mu$m and Pa$\beta$ lines. Instead, we follow \citet{Mouri2000} and translate the two [FeII]1.26$\mu$m/Pa$\beta$ limits proposed to distinguish between SB, AGN, and LINER excitation into values for the [FeII]1.64$\mu$m/Br$\gamma$ ratio. In Fig. \ref{fig:BPT}, we therefore plot [FeII]1.64$\mu$m/Br$\gamma$ and H$_2$/Br$\gamma$ for each of our nuclear and off-nuclear apertures, with correspondingly adjusted limits.

The line ratios for the nuclear (off-nuclear) apertures are represented by filled (open) circles. We only plot those apertures in which all three emission lines were detected with A/N$>3$.

The possible presence of reddening in our sources adds an additional uncertainty to the plotted [FeII]/Br$\gamma$ values. As the intrinsic reddening in each source cannot be determined from these long-slit spectra alone, we utilize NED\footnote{http://ned.ipac.caltech.edu/} and \citet{Ho1997} to estimate E(B-V) for several of our sources. In all those cases, E(B-V) does not exceed 1\,mag, more often it is $<$0.4mag. Applying the \citet{Calzetti1994} extinction law, this implies a 40\% (0.2dex) and 17\% (0.08dex) decrease in the observed [FeII]1.64$\mu$m/Br$\gamma$ ratio, respectively. The foreground reddening is minimal (E(B-V)$<$0.1mag) in all our sources, due to the general position away from the Galactic plane. Reddening is not an issue for the H$_2$/Br$\gamma$ values, since the wavelength of the H$_2$ transition lies again close to the Br$\gamma$ wavelength.

Most points in Fig. \ref{fig:BPT} occupy the AGN domain, although some (both nuclear and off-nuclear) fall into the SB and LINER regions. The range of values displayed here agrees with e.g. \citet{2010MNRAS.404..166R} or \citet{2011MNRAS.416..493R}. There is no discernible difference between nuclear and off-nuclear environments, which suggests that the influence of the AGN systematically extends over a large range. This would be in agreement with the models of \citet{2012MNRAS.422..252D}, who claim the dominant role of X-rays for gas excitation throughout the central region.

Histograms of the [FeII]/Br$\gamma$ and H$_2$/Br$\gamma$ ratios for the nuclear (black, dashed) and off-nuclear (grey, solid) environments are also included in the figure. Again, no significant difference can be observed, except a low-end tail of `nuclear' regions in the H$_2$/Br$\gamma$ distribution (Mrk\,3, NGC\,2992, NGC\,4051, and NGC\,4151). Those galaxies may contain less (molecular) gas overall, the molecular gas may be heated less efficiently by the AGN, or the gas temperature may be low(er). From the line-to-continuum ratios in Fig. \ref{fig:mosiacfrac}, a minimum in the H$_2$ density towards the nucleus is not readily apparent, although increases in Br$\gamma$ density are. H$_2$ emission arises in regions where H is pre-dominantly neutral, while Br$\gamma$ emission arises from the more ionized shells of such regions \citep[e.g.][]{1997ARA&A..35..179H}. A larger ionized volume closer to the nucleus, due to X-ray emission from the AGN, would thus explain these observations.

From Fig. \ref{fig:BPT} it further appears that the [FeII]/Br$\gamma$ and H$_2$/Br$\gamma$ ratios are correlated. This correlation is mostly driven by the low H$_{2}$/Br$\gamma$ nuclear points. When they are included, the Pearson {\it r} correlation coefficient is 0.50 (P$\cong$0.001), without them {\it r} falls to 0.35 (P$\cong$0.05). This is similar to the correlation observed by \citet{2013MNRAS.430.2002R} in a sample of 65 nuclear spectra of star forming, AGN, and LINER galaxies.

We also tested whether the similarity of the [FeII]/Br$\gamma$ and H$_2$/Br$\gamma$ ratio profiles (as inferred from their distributions, as shown in Fig. \ref{fig:mosiacratio}) might be a distinguishing factor in the NIR diagnostic diagram. As mentioned before, similar radial profiles indicate a single dominant excitation source within that radial range, while different profiles point to a combination of excitation sources. The last column of Table \ref{tab:fluxes} indicates our judgement whether the line ratio profiles are different (`d') or similar (`s'). Interestingly, the three nuclei with different line ratio profiles (NGC\,1320, NGC\,2273, and NGC\,3081) all are Seyfert 2 galaxies. The off-nuclear apertures with different profiles, on the other hand, are located in both Seyfert 1 and 2 galaxies.

\begin{figure*}
\begin{tabular}{c c}
\put(0,0){\includegraphics[width=8.6cm]{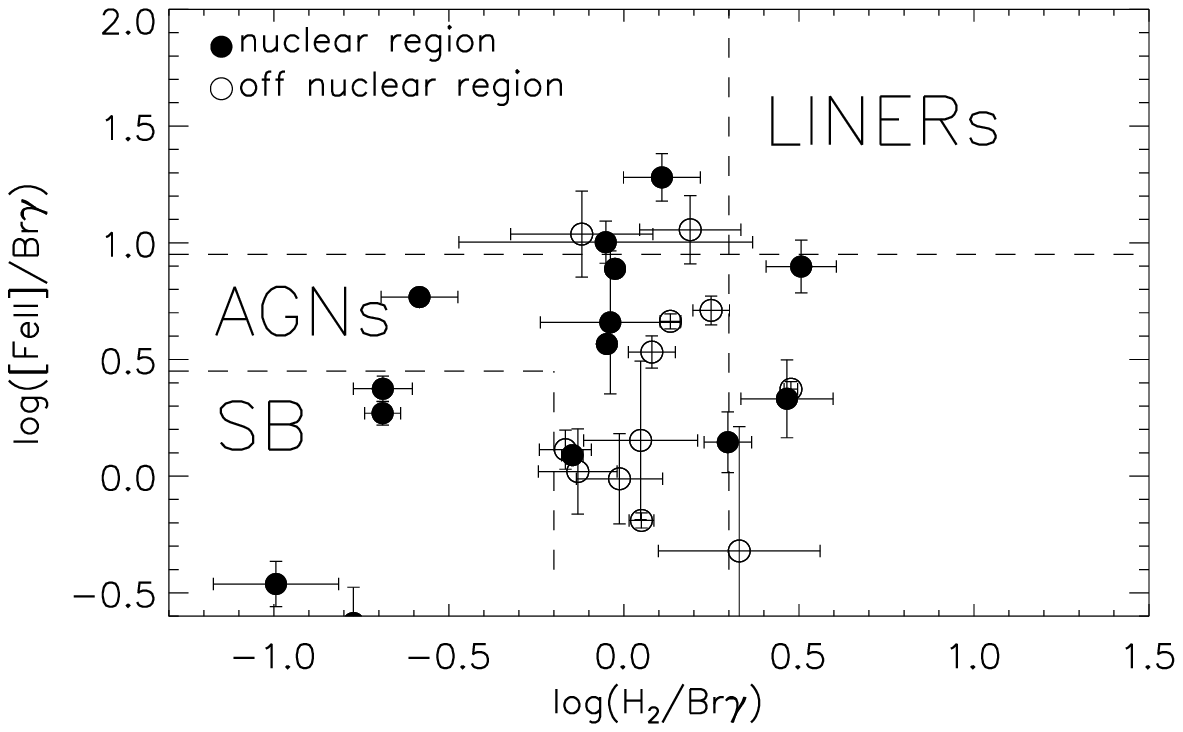}}
\put(3.5, 5.5){Similar radial variation}&
\put(8.8,0){\includegraphics[width=8.6cm]{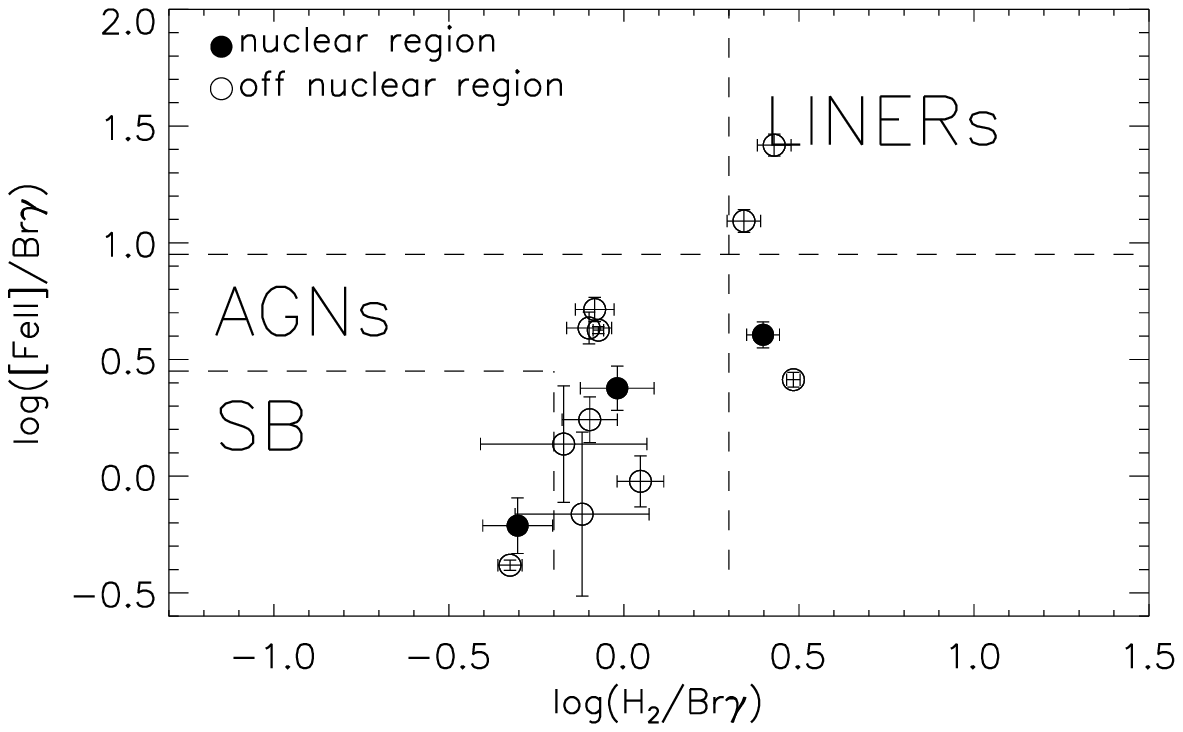}}
\put(12.7, 5.5){Different radial variation}\\
\end{tabular}
\caption{NIR diagnostic split by comparing the radial distribution of the [FeII]/Br$\gamma$ and H$_2$/Br$\gamma$ ratios within each aperture; similar radial distributions between the two ratios ({\it left}), and varying distributions ({\it right}). Similar here does not mean similar values, but only shape of the radial variations.}\label{fig:BPT_ratio}
\end{figure*}

In Fig. \ref{fig:BPT_ratio} the NIR diagnostic diagram is reproduced, now split along this division. It appears that the line ratios are more strongly correlated for apertures with different line ratio profiles. The spread of data-points over the SB, AGN, and LINER domains however remains in both groups. Particularly, in the `similar radial variation' group there is no clearer distribution of data-points in either the SB or AGN domain, than in the `different radial variation' group.

\section{Kinematics of gas and stars}
Kinematic information, which is a by-product of the emission line fitting with pPXF and GANDALF, is another useful diagnostic to better understand the physics of the ISM. 

\begin{figure*}
\centering
\includegraphics[]{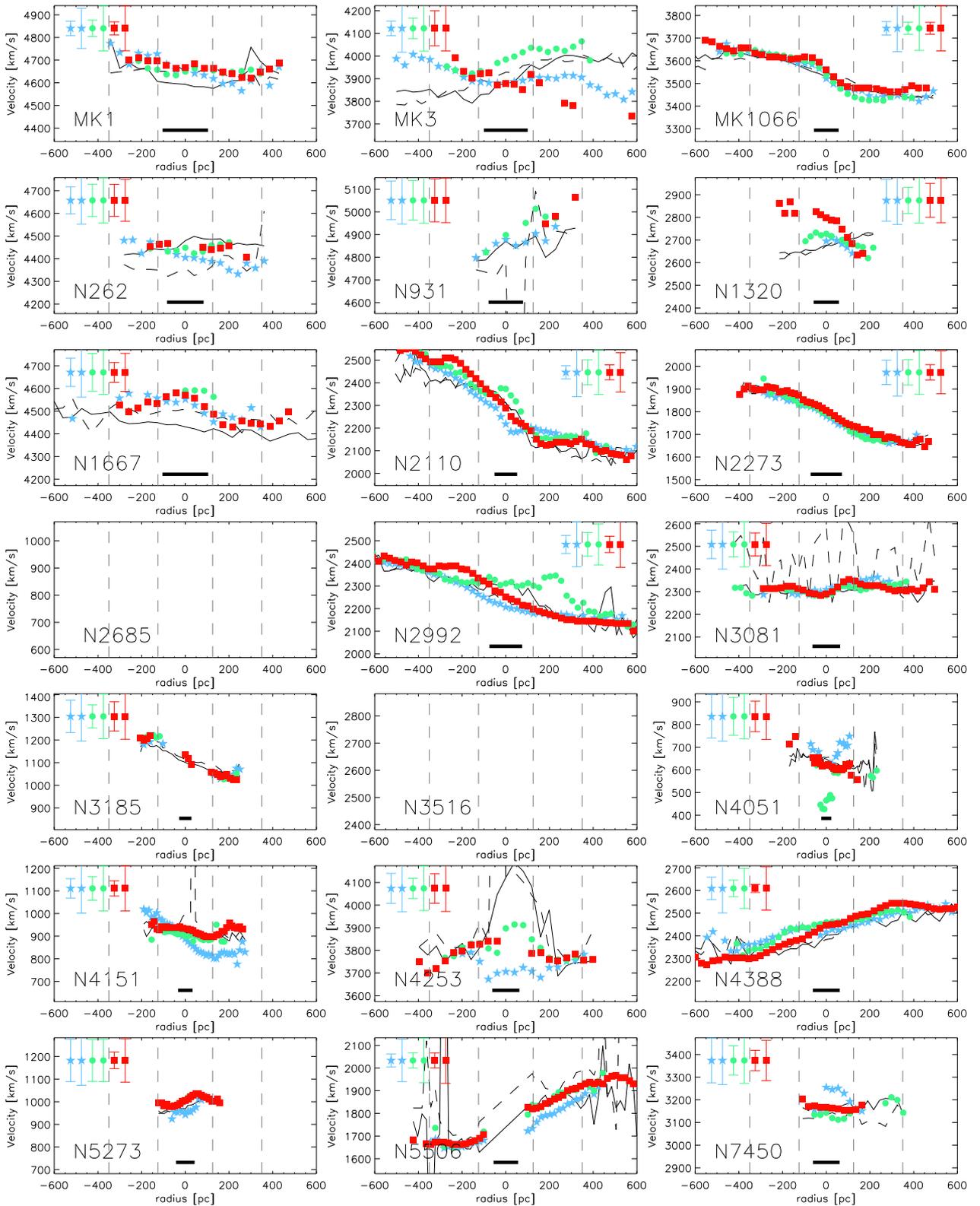}
\caption{The velocities derived from the stellar continuum and from the [FeII] (blue stars), Br$\gamma$ (green dots), and H$_2$ (red squares) emission lines by pPXF + GANDALF. The stellar H(/K) band is plotted as the solid(/dashed) line. Displayed points each have an S/N\,$>$\,3. The seeing (and nuclear aperture) is indicated with the black bar. The off-nuclear apertures are delimited by the vertical dashed lines at (+/-)375\,pc and (+/-)125\,pc. The minimum (20\,km/s) and maximum (40\,km/s) 1$\sigma$ errors for each emission line are shown in the upper right or left corner. {\it A color version of this figure is available in the online journal.}}\label{fig:mosiackin}
\end{figure*}

In Fig. \ref{fig:mosiackin} the velocities of the three lines are plotted as a function of radial distance from the nucleus, together with the best-fit stellar velocity in both H- and K-band. The NIRSPEC slit was positioned along the major axis of all target galaxies. In several sources, rotation-like curves are observed in the emission lines. We note that at first glance no distinction can be made between rotation and in/outflow when observing along a single slit. However, the gas velocity curves are similar to those of the stars, which define the galaxy disk. This is the case in Mrk\,1, Mrk\,1066, NGC\,262, NGC\,931, NGC\,1667, NGC\,2110, NGC\,2273, NGC\,3081, NGC\,3185, NGC\,4388, NGC\,5273, and NGC\,5506. In these galaxies, all three emission lines display a velocity curve that follows that of the stars and is reminiscent of rotation, thus strongly implying that the line-emitting gas is located in the disk of the galaxy and rotating with the disk. 

Other galaxies (Mrk\,3, NGC\,1320, NGC\,2992, NGC\,4051, NGC\,4151, NGC\,4253, and NGC\,7450) display velocity curves different from the stellar velocity curve in one or more emission lines, which may be a signature of non-rotational motion, either in- or outflows or other non-planar motion of the gas. Although by itself it provides no insight into the dominant gas excitation mechanism, we find that most sources with irregular (i.e. mutually varying, and not following the stellar velocity curve) gas dynamics have low H$_2$/Br$\gamma$ ratios in their nuclei, and thus fall into the SB domain of the NIR diagnostic diagram. This indicates a possible correlation between irregular gas dynamics in the central kiloparsec and a decreased amount of molecular gas at the nucleus proper.

\begin{figure*}
\centering
\includegraphics[]{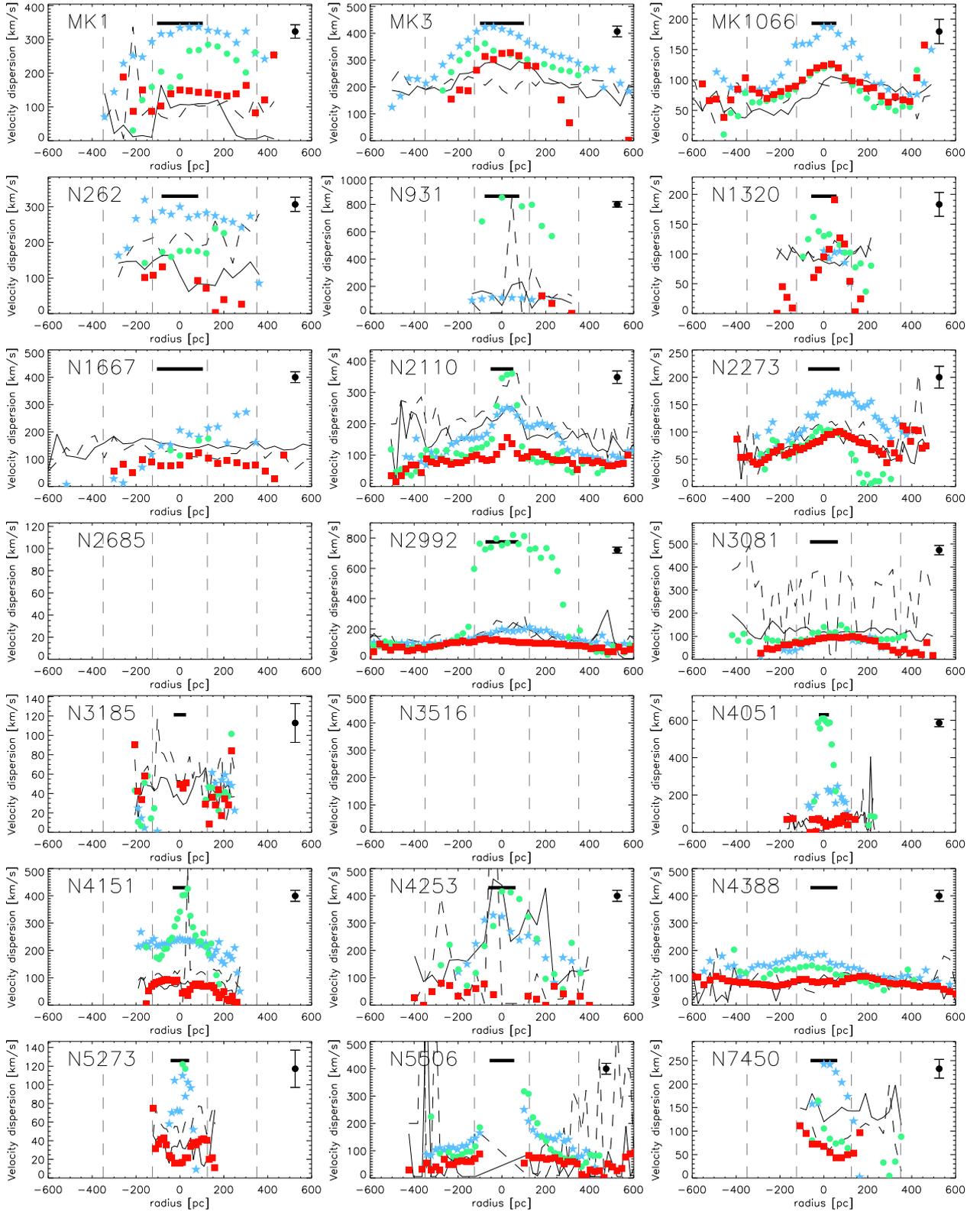}
\caption{The velocity dispersion derived from the stellar continuum and the [FeII] (blue stars), Br$\gamma$ (green dots) and H$_2$ (red squares) emission lines by pPXF + GANDALF. The stellar H(/K) band is plotted as the solid(/dashed) line. Displayed points each have an S/N\,$>$\,3. The seeing (and nuclear aperture) is indicated with the black bar. The off-nuclear apertures are delimited by the vertical dashed lines at (+/-)375\,pc and (+/-)125\,pc. The 1$\sigma$ (20\,km/s) uniform error is shown in the upper right corner. {\it A color version of this figure is available in the online journal.}}\label{fig:mosiacsig}
\end{figure*}

The velocity dispersion profiles of the three emission lines are shown in Fig. \ref{fig:mosiacsig}. The velocity dispersion was left a free parameter during the spectral fits. The velocity dispersion covers a broad range of values, and its profiles show a variety of distinct shapes. Three broad categories of radial profile shapes can be identified: centrally peaked, flat, and varying between gas species.

The radial dispersion profiles in Mrk\,3, Mrk\,1066, NGC\,2110, NGC\,2273, NGC\,4051, NGC\,4151, NGC\,4253, NGC\,5273, and NGC\,5506 have a central peak, which indicates a higher kinetic gas temperature in the nucleus. Either the ISM is more turbulent (possibly caused by a jet- or wind-driven outflow), and/or more intensely heated by the incident radiation field. In either case, a peaked dispersion distribution is a strong indication for the AGN governing the physical state of the gas at the nucleus. 
The dispersion peak is most often present in [FeII] (indicating dust destruction close to the nucleus), and/or in Br$\gamma$ (indicating a strong radiation field), while the H$_2$ line shows the least-pronounced central rise. 

Examples for flat velocity dispersion profiles are NGC\,3081, NGC\,3185, and NGC\,4388, which show no significant change over up to 1.2\,kpc in diameter. The velocity dispersion in all three sources is of order $\leq$100\,km/s, well within the range observed 
for star forming regions, which is consistent with star formation being the dominant the gas excitation mechanism.

Finally, the dispersion profiles in Mrk\,1, NGC\,262, NGC\,931, NGC\,1320, NGC\,1667, NGC\,2992, and NGC\,7450 show marked differences between the three emission lines, with variations of 100\,km/s or more at a given radius. This indicates that the ionized atomic and molecular gas is responding differently to the central AGN in these sources, perhaps due to the presences of asymmetric energy input (a jet or shocks) together with a stratified medium, e.g. a thin molecular disk in a thick ionized medium.

\section{Discussion}
The analyses presented in the previous sections have highlighted the diversity and complexity of gas excitation in (active) galactic nuclei. It is therefore challenging to draw general conclusions, especially given the relatively small size of the galaxy sample described in this work. Nevertheless, we have identified three main trends:  \\
i) galaxies in which the [FeII]/Br$\gamma$ and H$_2$/Br$\gamma$ line ratios follow different radial profiles within their nuclear aperture (indicated by `d' in the last column of Table \ref{tab:fluxes}), have peaked [FeII] line-to-continuum ratios and have Seyfert 2 nuclei. \\
ii) galaxies that fall into the `SB' domain of Fig. \ref{fig:BPT} due to their low H$_2$/Br$\gamma$ ratio in the nucleus, have emission lines that show centrally peaked velocity dispersions, and velocities that do not follow the stellar rotation curve. \\
 iii) galaxies in which the gas velocity dispersion shows little variation with radius (see Fig. \ref{fig:mosiacsig}), show gas that co-rotates with the stellar disk, and have Seyfert 2 nuclei.

Relations i) and ii) both deal with variations in the nuclear region, which are most likely linked to the presence of an AGN. Due to their higher energy, X-ray photons produced by the AGN can penetrate further into the ISM than UV-photons. Given that relation i) only holds for Seyfert 2 nuclei, for which - according to the standard AGN paradigm - the line-of-sight crosses the dust torus, a plausible explanation for the increase in the [FeII] line-to-continuum ratio are AGN-produced X-rays which penetrate the torus, and liberate Fe atoms from its dust. Similarly, the lower H2/Br$\gamma$ ratios in relation ii) are most likely linked to an increase in the Br$\gamma$ emission. A higher Br$\gamma$ flux can also be explained by X-ray heating from the AGN. Alternatively, it can be caused by UV radiation from a (circum)nuclear starburst. Relation iii), on the other hand, shows that even if the gas is heated by an AGN, this does not necessarily lead to non-rotational gas motions.

A number of recent studies using the Gemini Near-Infrared Integral Field Spectrpgraph (NIFS) can be used to compare to our results. Several of the objects studied with this instrument are also present in our sample, such as NGC\,4051 \citep{2008MNRAS.385.1129R}, Mrk\,1066 \citep{2010MNRAS.404..166R}, and NGC\,4151 \citep{Storchi2009}. All these papers present 2D distributions of the same emission lines as in our study, generally spanning the central 500\,pc in radius. In the galaxies just mentioned, as well as in others (NGC\,1068, \citet{2009ApJ...691..749M}, NGC\,7582, \citet{2009MNRAS.393..783R}, Mrk\,1157, \citet{2011MNRAS.416..493R} and, Mrk\,79, \citet{2013MNRAS.430.2249R}), two features are consistently seen. 

The first feature is that the heating of both H$_2$ and [FeII] is dominated by AGN X-rays, with a secondary contribution of shocks to the [FeII] emission. As mentioned above, our results show the spatially varying influence of the AGN, in either mechanical or thermal form, on the gas heating in the central kiloparsec. The dominant role of X-rays in (some) Seyfert galaxies may also be inferred from the intense soft X-ray emission found in the central 3\,kpc of NGC\,4151 by \citet{2011ApJ...729...75W}, or \citet{2012ApJ...755...57H}, who find that AGN-based heating mechanisms must play an important role in the central kiloparsec in NGC\,1068.  

The second feature generally seen is that the H$_2$ emitting gas is located in the disk of the galaxy, while the [FeII] emitting gas is connected with gas outflow (due to the AGN). This is primarily a kinematic argument, with which our kinematics generally agree. The velocity dispersion in H$_2$ rarely exceeds 100\,km/s, which is generally true for disk gas. However, while [FeII] and Br$\gamma$ lines do show higher velocity dispersion values, we don't always see outflow-like signatures in the velocity curves of the probed [FeII] and Br$\gamma$ lines.

\subsection{Aperture size}
The previous discussion is dominated by the combination of aperture integrated values and radial distributions. Therefore, we also wish to spend a few thoughts on the choice of apertures. It will be evident that the choice of range within which spectra are integrated strongly influences the measured emission line ratios, and thus the inferred position in the NIR diagnostic diagram in Fig. \ref{fig:BPT}. 

The variety of [FeII]/Br$\gamma$ and H$_2$/Br$\gamma$ line ratio values at different spatial positions is apparent from Fig. \ref{fig:mosiacratio}. When integrating over any extended apertures, these local variations will of course be averaged out. This implies that the aperture size used to extract the spectrum will have a profound effect on the measured line ratios. In particular, the drop of the H$_2$/Br$\gamma$ ratios in the nucleus can only be sampled with seeing limited apertures as we used for the nuclear apertures in this work.

For the off-nuclear apertures we were more constrained by arguments of spatial uniformity across the sample, than seeing limits. Interestingly, we find that spectral apertures with the more complex line ratio profiles appear to have the tightest correlation in the adapted NIR diagnostic diagram. Does this reflect the disjoint origins of the [FeII] (mechanical) and H$_2$ (thermal) excitation mechanisms. Which only correlate when both are sufficiently sampled within an aperture?

\section{Summary}
Taking advantage of the large aperture of the Keck telescope and the excellent atmospheric conditions at Mauna Kea, we have observed a sample of 21 nearby Seyfert galaxies in the H- and K-bands. The high sensitivity of the NIRSPEC instrument allows us to use a small slit width, which makes it possible to detect key emission lines of the ISM, and to separate the nuclear and off-nuclear environments in each galaxy. The spatial extent probed by these observations allowed us to systematically compare the physical state of the ISM across the central kiloparsec, including the immediate vicinity of the Seyfert nucleus. We analyzed the gas excitation across the galaxy sample using gas and stellar rotation curves, velocity dispersion distributions, and line-to-continuum ratios, all out to 600\,pc radius. Additionally extracting spectra from both nuclear and off-nuclear apertures, we have investigated the dominant excitation mechanism (starburst, AGN or LINER) acting on the ionized and molecular gas by means of a NIR diagnostic diagram based on the [FeII]1.64$\mu$m/Br$\gamma$ vs. H$_2$/Br$\gamma$ ratios. We find no clear distinction between nuclear and off-nuclear environments in our adopted NIR diagnostic diagram. The influence of AGN on gas excitation appears to reach far into the central kiloparsec of each galaxy, besides the expected local star formation.

Our investigations show the complexity that is possible in the interplay between excitation arising from the AGN as well as local star formation. However, the influence of X-ray emission from the AGN does appear to play a significant role in all sources. By invoking the standard AGN-paradigm, some further interpretation of individual results may be reached, although we are limited by small number statistics.

One interesting relation we found was that three Seyfert nuclei in our sample would be considered SB nuclei in one of our main analyses, due to the measured low H$_2$/Br$\gamma$ emission line ratio. A drop in this ratio around the nucleus, which is seen to a greater or lesser extent in many sources in our sample, can only be observed in well-resolved spectroscopic data, or seeing-limited apertures. 

The reduced and calibrated long-slit spectra as well as the fitting results will be made available online.

\begin{acknowledgements}
We thank the anonymous referee for valuable comments that improved this paper. The authors wish to recognize and acknowledge the very significant cultural role and reverence that the summit of Mauna Kea has always had within the indigenous Hawaiian community.  We are most fortunate to have the opportunity to conduct observations from this mountain. We wish to thank Jesus Falcon-Barroso for help with the spectral fitting. Our thanks also go to Ariane Lan\c{c}on for generating synthetic spectra from her models specifically at our spectral resolution. TvdL would like to acknowledge P.D. Barthel, who once upon a time wanted to see these data sets published and got her and ES talking. Part of this work was done during an 2011-2012 IPAC visiting graduate fellowship. The Digitized Sky Survey was produced at the Space Telescope Science Institute under U.S. Government grant NAG W-2166. The images of these surveys are based on photographic data obtained using the Oschin Schmidt Telescope on Palomar Mountain and the UK Schmidt Telescope. The plates were processed into the present compressed digital form with the permission of these institutions.
\end{acknowledgements}
\bibliographystyle{aa}
\bibliography{SeypaperI}

\clearpage

\begin{table*}
\setlength{\tabcolsep}{0.050in} 
\caption{Key emission line integrated fluxes and ratios.}\label{tab:fluxes}
\centering
\begin{tabular}{l c c c c c c}
\hline\hline
Source & [FeII] 1.64$\mu$m & H$_2$(1-0)S(1) & Br$\gamma$ & [FeII]/Br$\gamma$ & H$_2$/Br$\gamma$ & d/s \\
\hline
Mrk\,1 & 24.9$\pm $0.7 & 3.04$\pm $0.08 & 3.22$\pm $0.13 & 7.7$\pm $0.05 & 0.94$\pm $0.006 & s\\
\dots & 4.7$\pm $0.2 & 0.46$\pm $0.02 & 0.42$\pm $0.03 & 11.2$\pm $0.1 & 1.09$\pm $0.01 & s\\
\dots & 6.6$\pm $0.3 & 0.61$\pm $0.03 & 0.49$\pm $0.04 & 13.5$\pm $0.2 & 1.24$\pm $0.01 & s\\
\hline
Mrk\,3 & 363$\pm $11 & 16$\pm $4 & 62$\pm $4 & 5.85$\pm $0.05 & 0.26$\pm $0.02 & s\\
\dots & 177$\pm $6 & 3.5$\pm $1.0 & 14.5$\pm $1.5 & 12.2$\pm $0.1 & 0.24$\pm $0.02 & s\\
\dots & 93$\pm $5 & 5.4$\pm $1.7 & 24.1$\pm $1.9 & 3.84$\pm $0.05 & 0.22$\pm $0.02 & s\\
\hline
Mrk\,1066 & 7.1$\pm $0.3 & 1.72$\pm $0.05 & 1.93$\pm $0.05 & 3.68$\pm $0.04 & 0.89$\pm $0.007 & s\\
\dots & 4.08$\pm $0.11 & 0.82$\pm $0.02 & 0.97$\pm $0.02 & 4.22$\pm $0.03 & 0.84$\pm $0.006 & d\\
\dots & 1.39$\pm $0.07 & 0.410$\pm $0.017 & 0.302$\pm $0.016 & 4.6$\pm $0.06 & 1.36$\pm $0.01 & s\\
\hline
NGC\,262 & 4.8$\pm $0.5 &   & 0.48$\pm $0.09 & 10.1$\pm $0.3 &   & s\\
\dots & 0.89$\pm $0.12 & 0.055$\pm $0.013 &   &   &   & s\\
\dots & 0.80$\pm $0.12 & 0.036$\pm $0.006 & 0.061$\pm $0.017 & 13$\pm $1.0 & 0.59$\pm $0.03 & s\\
\hline
NGC\,1320 & 0.58$\pm $0.15 & 0.47$\pm $0.10 & 0.94$\pm $0.10 & 0.61$\pm $0.04 & 0.50$\pm $0.03 & d\\
\dots &   & 0.12$\pm $0.02 &   &   &   \\
\dots &   &   & 0.061$\pm $0.019 &   &  \\
\hline
NGC\,1667 & 1.46$\pm $0.19 & 0.59$\pm $0.04 & 0.18$\pm $0.04 & 7.9$\pm $0.3 & 3.21$\pm $0.05 & s\\
\dots & 0.36$\pm $0.06 & 0.251$\pm $0.017 &   &   &   \\
\dots & 0.74$\pm $0.12 & 0.228$\pm $0.013 &   &   &  \\
\hline
NGC\,2273 & 1.99$\pm $0.15 & 1.23$\pm $0.04 & 0.49$\pm $0.05 & 4.0$\pm $0.08 & 2.50$\pm $0.02 & d\\
\dots & 0.85$\pm $0.08 & 0.30$\pm $0.03 & 0.25$\pm $0.03 & 3.4$\pm $0.08 & 1.20$\pm $0.03 & s\\
\dots & 1.17$\pm $0.12 & 0.40$\pm $0.03 & 0.23$\pm $0.02 & 5.1$\pm $0.1 & 1.77$\pm $0.03 & s\\
\hline
NGC\,2685 &   &   &   &   &  \\
\dots &   &   &   &   &  \\
\dots &   &   &   &   &  \\
\hline
NGC\,3081 & 1.47$\pm $0.18 & 0.59$\pm $0.10 & 0.62$\pm $0.11 & 2.4$\pm $0.07 & 0.96$\pm $0.04 & d\\
\dots & 0.24$\pm $0.05 & 0.28$\pm $0.03 & 0.25$\pm $0.03 & 1.0$\pm $0.05 & 1.11$\pm $0.03 & d\\
\dots & 0.41$\pm $0.08 & 0.19$\pm $0.02 & 0.24$\pm $0.03 & 1.7$\pm $0.08 & 0.80$\pm $0.03 & d\\
\hline
NGC\,3185 &   & 0.056$\pm $0.014 &   &   &   \\
\dots &   & 0.040$\pm $0.008 & 0.055$\pm $0.009 &   & 0.74$\pm $0.04 & s\\
\dots & 0.114$\pm $0.019 & 0.060$\pm $0.008 & 0.088$\pm $0.009 & 1.3$\pm $0.05 & 0.68$\pm $0.02 & s\\
\hline
NGC\,4388 & 5.1$\pm $0.4 & 2.98$\pm $0.16 & 4.18$\pm $0.19 & 1.23$\pm $0.02 & 0.71$\pm $0.009 & s\\
\dots & 2.87$\pm $0.17 & 3.38$\pm $0.03 & 1.11$\pm $0.05 & 2.59$\pm $0.04 & 3.05$\pm $0.008 & d\\
\dots & 2.37$\pm $0.15 & 3.01$\pm $0.05 & 1.00$\pm $0.04 & 2.36$\pm $0.04 & 3.00$\pm $0.01 & s\\
\hline
NGC\,931 & 0.8$\pm $0.2 &   & 5.4$\pm $1.5 & 0.15$\pm $0.010 &  \\
\dots &   &   &   &   &  \\
\dots & 0.11$\pm $0.03 & 0.14$\pm $0.05 & 0.80$\pm $0.19 & 0.14$\pm $0.009 & 0.18$\pm $0.01 & s\\
\hline
NGC\,2110 & 14.9$\pm $0.5 & 1.01$\pm $0.10 & 0.78$\pm $0.18 & 19.1$\pm $0.1 & 1.28$\pm $0.03 & s\\
\dots & 3.23$\pm $0.14 & 0.57$\pm $0.02 & 0.26$\pm $0.03 & 12.4$\pm $0.1 & 2.20$\pm $0.02 & d\\
\dots & 5.86$\pm $0.17 & 0.60$\pm $0.02 & 0.22$\pm $0.02 & 26.2$\pm $0.2 & 2.69$\pm $0.03 & d\\
\hline
NGC\,2992 & 17.3$\pm $1.3 & 1.90$\pm $0.14 & 9.3$\pm $0.8 & 1.86$\pm $0.04 & 0.204$\pm $0.004 & s\\
\dots & 5.8$\pm $0.3 & 0.92$\pm $0.06 & 1.12$\pm $0.12 & 5.2$\pm $0.07 & 0.82$\pm $0.01 & d\\
\dots & 6.0$\pm $0.4 & 1.11$\pm $0.05 & 1.39$\pm $0.20 & 4.3$\pm $0.07 & 0.80$\pm $0.008 & d\\
\hline
NGC\,3516 &   &   & 18$\pm $4 &   &  \\
\dots &   &   &   &   &  \\
\dots &   &   &   &   &  \\
\hline
NGC\,4051 & 1.2$\pm $0.3 & 0.9$\pm $0.2 & 5.2$\pm $1.2 & 0.23$\pm $0.02 & 0.17$\pm $0.010 & s\\
\dots &   & 0.074$\pm $0.019 &   &   &  \\
\dots &   & 0.073$\pm $0.019 & 0.065$\pm $0.018 &   & 1.1$\pm $0.07\\
\hline
NGC\,4151 & 22.5$\pm $1.3 & 1.9$\pm $0.3 & 9.5$\pm $1.1 & 2.37$\pm $0.03 & 0.20$\pm $0.008 & s\\
\dots & 1.3$\pm $0.2 & 0.09$\pm $0.02 &   &   &   & s\\
\dots & 1.40$\pm $0.18 & 0.19$\pm $0.02 & 0.12$\pm $0.04 & 11.4$\pm $0.4 & 1.55$\pm $0.05 & s\\
\hline
NGC\,4253 & 9.9$\pm $1.6 &   & 29$\pm $5 & 0.35$\pm $0.01 &   & s\\
\dots & 1.4$\pm $0.3 & 0.83$\pm $0.15 & 2.2$\pm $0.3 & 0.65$\pm $0.03 & 0.37$\pm $0.02 & d\\
\dots & 0.98$\pm $0.13 & 0.42$\pm $0.07 & 1.54$\pm $0.12 & 0.64$\pm $0.02 & 0.28$\pm $0.01 & d\\
\hline
NGC\,5273 & 0.33$\pm $0.08 & 0.45$\pm $0.03 & 0.15$\pm $0.05 & 2.1$\pm $0.1 & 2.92$\pm $0.04\\
\dots &   &   &   &   &  \\
\dots &   & 0.030$\pm $0.009 &   &   &  \\
\hline
NGC\,5506 & 18$\pm $2 &   &   &   &  \\
\dots & 3.36$\pm $0.10 & 3.8$\pm $0.3 & 8.1$\pm $0.3 & 0.415$\pm $0.003 & 0.47$\pm $0.008 & d\\
\dots & 5.04$\pm $0.15 & 8.8$\pm $0.4 & 7.8$\pm $0.5 & 0.646$\pm $0.005 & 1.12$\pm $0.01 & s\\
\hline
NGC\,7450 & 0.56$\pm $0.15 & 0.79$\pm $0.05 & 0.40$\pm $0.06 & 1.4$\pm $0.09 & 1.98$\pm $0.03 & s\\
\dots &   & 0.070$\pm $0.019 &   &   &  \\
\dots &   & 0.074$\pm $0.015 & 0.077$\pm $0.015 &   & 0.97$\pm $0.05\\
\hline
\hline
\end{tabular}
\tablefoot{Fluxes for [FeII], H$_2$(1-0), and Br$\gamma$ emission lines, as well as their [FeII]/Br$\gamma$ and H$_2$/Br$\gamma$ ratios for the nuclear and off-nuclear integrated apertures. All fluxes are in units of $10^{-15}$\,erg\,s$^{-1}$\,cm$^{-2}$. Top entry for each source gives data for nuclear aperture, bottom two for off-nuclear apertures. Where no significant line has been detected, the entry has been left empty. In the final column `d' and `s' indicate different or similar radial variation, as used in Fig. \ref{fig:BPT_ratio}.}
\end{table*}

\setlength{\unitlength}{1cm}

\clearpage
\onecolumn
\begin{appendix}
\section{Available data products}
The following data products (FITS-files) are made available. 

The {\it calibrated long-slit spectra} contain the long-slit spectra, fully calibrated as described in Section \ref{sec:datareduc}. All pertinent information concerning the observations (position, observation date, airmass, wavelengths, flux conversion factor etc.) are contained in the standard FITS header. In the second layer of this file, the ``correction spectrum'' is appended. 

The {\it fitting results} contain the results of the pPXF + GANDALF fits (Section \ref{sec:GANDALF}). This multi-layered file contains 1) the input spectrum (rebinned to ln, multiplied by 1e13 for computational reasons), 2) the best fitting model, 3) the best fitting emission line spectrum, 4) the best fitting spectrum with emission lines removed, 5) the derived stellar and gaseous parameters, 6) normalized weights assigned to each stellar template, and 7) the optimal combination of stellar templates. 

The format of layer 5) is a 2-dimensional matrix with 101 columns (i.e. for each position in the long-slit spectrum). Each row contains the stellar kinematics; V, $\sigma$, h3, h4, h5, h6, and the gaseous kinematics; flux, amplitude, V, $\sigma$, Amplitude-over-Noise of each line (5/13 emission lines in the H/K band, see Table \ref{tab:lines}) for each position in the slit. The outer positions, where the S/N of the observations was extremely poor, were set to zero to reduce computation time. 

This file structure is directly generated by pPXF and GANDALF, see the README file distributed with the GANDALF code for more information (http://star-www.herts.ac.uk/$\sim$sarzi/PaperV\_nutshell/PaperV\_nutshell.html).

Further, {\it nuclear / off-nuclear integrated spectra} are available in the same formats. These data offer a simple comparison between nuclear and off-nuclear regions in the full long-slit spectra. Calibrated spectra are given for a nuclear and two off-nuclear (symmetric about the nucleus) regions. The positions used to generate these summed spectra are indicated in the FITS-header. The pPXF + GANDALF fitting result contains a further fourth spectrum, which is the sum of the two off-nuclear region spectra. 
\end{appendix}

\end{document}